# GENERATION OF NARROW BEAMS OF SUPER HIGH-ENERGY GAMMA QUANTA IN THE RESONANT INVERSE COMPTON EFFECT IN THE FIELD OF A STRONG X-RAY WAVE

Sergei P. Roshchupkin, Sergey B. Makarov

Peter the Great St. Petersburg Polytechnic University,
195251, St-Petersburg, Russian Federation, Russia
serg9rsp@gmail.com; makarov@cee.spbstu.ru

The article presents a theoretical study of Oleinik resonances in the process of scattering a gamma quantum by an ultrarelativistic electron in the field of a strong electromagnetic wave with intensities up to $10^{27}$ Wcm$^{-2}$. The resonant kinematics for three possible resonant reaction channels in a strong external field has been studied in detail. It is shown that under resonant conditions, the scattering channels of the reaction effectively split into two first-order processes according to a fine structure constant such as the external field-stimulated Compton effect. And the annihilation channel of the reaction effectively decays into direct and reverse the external field-stimulated Breit-Wheeler processes. In the absence of interference of the reaction channels, a resonant differential cross section was obtained in a strong external electromagnetic field. The significant dependence of the resonant energy of final particles and resonant cross sections on the outgoing angles of the final gamma quantum, the number of absorbed and emitted photons of the wave, as well as the characteristic quantum parameters of the problem is shown. These quantum parameters are determined by the ratio of the initial particle energies to the characteristic energies of the Compton effect and the Breit-Wheeler process. An unambiguous relationship between the outgoing angles of final electrons and gamma quanta has been obtained, which qualitatively distinguishes the resonant process from the non-resonant one. The cases when the energy of the initial electrons significantly exceeds the energy of the initial gamma quanta have been studied. The conditions under which the energy of high-energy initial electrons is converted into the energy of final gamma quanta are obtained. At the same time, the resonant differential cross-section of such a process significantly (by several orders of magnitude) exceeds the corresponding non-resonant cross-section. This theoretical study predicts a number of new physical effects that may explain the high-energy fluxes of gamma quanta produced near neutron stars and magnetars.

## 1. Introduction

Currently, high-intensity laser systems are being intensively developed [1-9], as well as sources of high-energy particles, including high-energy gamma quanta [10-17]. All this contributes to the intensive development of quantum electrodynamics (QED) in strong electromagnetic fields (see, for example, reviews [18-27], monographs [28-31] and articles [32-87]). It is important to emphasize that QED processes of higher orders with respect to the fine structure constant in the laser field (QED processes modified by an external electromagnetic field) can proceed in both resonant and non-resonant ways. The so-called Oleinik resonances may occur here [32, 33] (see also articles [45-60]) due to the fact that lower-order processes are allowed in the electromagnetic field with respect to the fine structure constant (QED processes stimulated by an external electromagnetic field). It is important to emphasize that the resonant differential cross sections can significantly exceed the corresponding nonresonant differential cross sections [20, 24, 54-60].

The Compton effect stimulated by an external electromagnetic field (a first-order process by a fine structure constant) has been studied for a long time (see, for example, the review [19] and articles [1, 34-44]). The resonant Compton effect modified by an external electromagnetic field (a second-order process with respect to a fine structure constant) was previously studied in the field of a weak electromagnetic wave (see, for example, [33, 45-49]).

In this article, unlike the previous ones, we will study the resonant inverse Compton effect modified by the strong field of a plane circularly polarized wave for ultrarelativistic electron energies. In this case, the main parameter is the classical relativistically invariant parameter

$$\eta = \frac{eF\lambdabar}{mc^2} \gtrsim 1 \tag{1}$$

which is numerically equal to the ratio of the work of the field at the wavelength to the rest energy of the electron ($e$ and $m$ are the charge and mass of the electron, $F$ and $\lambdabar = c/\omega$ are the electric field strength and wavelength, and $\omega$ is the frequency of the wave). In addition, in this problem, for three resonant reaction channels (see Feynman diagrams in Figure 2), characteristic quantum parameters arise equal to the ratio of the energies of the initial particles to the characteristic energies of the process:

$$\varepsilon_{iC} = \frac{E_i}{\hbar\omega_C}, \quad \varepsilon'_{iC} = \frac{\omega_i}{\omega_C}, \quad \varepsilon'_{iBW} = \frac{\omega_i}{\omega_{BW}}. \tag{2}$$

Here $E_i$ and $\hbar\omega_i$ are the energies of the initial electrons and gamma quanta, as well as $\hbar\omega_C$ and $\hbar\omega_{BW}$ are the characteristic quantum energies of the Compton effect [56, 59] and the Breit-Wheeler process [54, 55, 58]:

$$\hbar\omega_C = \frac{(m_*c^2)^2}{4(\hbar\omega)\sin^2(\theta/2)} = \frac{(mc^2)^2(1+\eta^2)}{4(\hbar\omega)\sin^2(\theta/2)}, \quad \omega_{BW} = 4\omega_C. \tag{3}$$

Here $m_*$ is the effective mass of an electron in the field of a circularly polarized wave (12), $\theta$ is the angle between the momentum of the initial gamma quantum and the direction of propagation of the wave (22). Note that the characteristic energies (3) are inversely proportional to the frequency $(\omega)$ and directly proportional to the intensity of the external electromagnetic wave $(I \sim \eta^2)$, and also depend on the angle between the momenta of the wave and the initial particles. In this paper, it will be shown that the resonant energies of final electrons and gamma quanta, as well as the resonant differential cross sections, significantly depend on the magnitude of the quantum parameters $\varepsilon_{iC}, \varepsilon'_{iC}$ and $\varepsilon'_{iBW}$ (2).

Later in the article, the relativistic system of units is used: $\hbar = c = 1$.

## 2. The amplitude of the scattering of a gamma quantum by an electron in a strong light field

The process of scattering of a gamma quantum by an electron in an external electromagnetic field is a second-order process according to the fine structure constant and is described by two Feynman diagrams (see Fig. 1). Let's choose the 4-potential of a plane monochromatic circularly polarized electromagnetic wave propagating along the $z$ axis in the following form:

$$A(\phi) = \frac{F}{\omega}(e_x\cos\phi + \delta \cdot e_y\sin\phi), \quad \phi = kx = \omega(t-z). \tag{4}$$

Here $\delta = \pm 1$; $e_{x,y} = (0, \mathbf{e}_{x,y})$ and $k = (\omega, \mathbf{k})$ are 4-polarization vectors and 4-wave vector of the external field, $k^2 = 0, e_{x,y}^2 = -1, (ke_{x,y}) = 0$.

The wave functions of an electron are determined by the Volkov functions [80, 81], the intermediate states of an electron (positron) are given by the Green function in the field of a plane wave (4) [82, 83]. The amplitude of such a process after simple calculations can be represented in the following form (see, for example, [45-49, 85]):

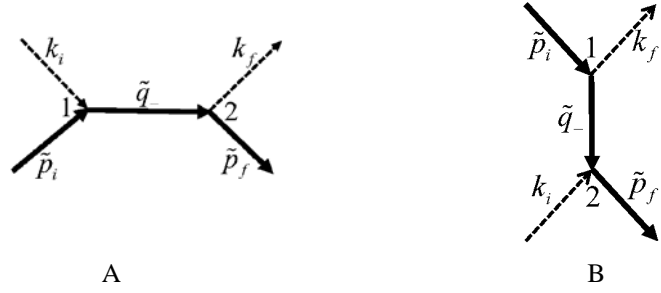

A            B

Fig. 1. Feynman diagrams of the scattering process of a gamma quantum on an electron in the field of a plane electromagnetic wave. The solid incoming and outgoing lines correspond to the Volkov functions of the electron in the initial and final states, the inner line corresponds to the Green function of the electron in the plane wave field. The dashed lines correspond to the wave functions of the gamma quantum in the initial and final states.

$$S_{fi} = \sum_{l=-\infty}^{\infty} S_l, \qquad (5)$$

where the partial amplitude with emission and absorption of $l$-photons of the wave has the following form:

$$S_l = -i\frac{(2\pi)^4 e^2}{\sqrt{\tilde{E}_i \tilde{E}_f \omega_i \omega_f}} \cdot \exp(i\varphi_{fi}) \cdot [\bar{u}_f M_l u_i] \cdot \delta^{(4)}[\tilde{p}_f + k_f - \tilde{p}_i - k_i - lk], \qquad (6)$$

It is indicated here:

$$l = r_2 - r_1, \qquad (7)$$

$$M_l = \varepsilon_\mu \varepsilon_\nu^{*} \sum_{r_2=-\infty}^{\infty} \left[ F_{r_2}^\nu(\tilde{p}_f, \tilde{q}_-) \cdot \left(\frac{\hat{q}_- + m}{\tilde{q}_-^2 - m_*^2}\right) F_{r_2-l}^\mu(\tilde{q}_-, \tilde{p}_i) + F_{r_2}^\mu(\tilde{p}_f, \tilde{q}_-)\left(\frac{\hat{q}_- + m}{\tilde{q}_-^2 - m_*^2}\right) F_{r_2-l}^\nu(\tilde{q}_-, \tilde{p}_i) \right]. \qquad (8)$$

In terms of (6)-(8) $\varepsilon_\mu$, $\varepsilon_\nu^{*}$ are 4- polarization vectors of the initial and final gamma quanta, $\varphi_{fi}$ is a phase independent of the summation indices, $u_i, \bar{u}_f$ are Dirac bispinors, $\tilde{p}_i = (\tilde{E}_i, \tilde{\mathbf{p}}_i)$ and $\tilde{p}_f = (\tilde{E}_f, \tilde{\mathbf{p}}_f)$ are 4-quasi-momenta of the initial and final electrons, $\tilde{q}_-$ is the 4-quasi-momentum of the intermediate electron, $m_*$ is the effective mass of the electron in the plane wave field [19]. At the same time, for channels A and B, you can write:

$$A: \quad \tilde{q}_- = \tilde{p}_i + k_i - r_1 k = \tilde{p}_f + k_f - r_2 k, \qquad (9)$$

$$B: \quad \tilde{q}_- = \tilde{p}_i - k_f + r_1 k = \tilde{p}_f - k_i + r_2 k \qquad (10)$$

$$\tilde{p}_{i,f} = p_{i,f} + \eta^2 \frac{m^2}{2(kp_{i,f})} k, \quad \tilde{q}_- = q_- + \eta^2 \frac{m^2}{2(kq_-)} k, \qquad (11)$$

$$\tilde{p}_{i,f}^2 = \tilde{q}_-^2 = m_*^2, \quad m_* = m\sqrt{1+\eta^2}. \qquad (12)$$

Here $k_{i,f} = \omega_{i,f}(1, \mathbf{n}_{i,f})$ are 4-the momentum of the initial and final gamma quantum, $p_{i,f} = (E_{i,f}, \mathbf{p}_{i,f})$ are 4-the momentum of the initial and final electrons. Expressions with a cap in the ratio (8) and further mean the scalar product of the corresponding 4-vector on the Dirac gamma matrices: $\gamma^\mu = (\gamma^0, \boldsymbol{\gamma}), \mu = 0,1,2,3$. For example, $\hat{\tilde{p}}_i = \tilde{p}_{i\mu}\gamma^\mu = \tilde{p}_{i0}\gamma^0 - \tilde{\mathbf{p}}_i\boldsymbol{\gamma}$. The amplitudes $F_{r'-l}^{\nu(\mu)}(\tilde{q}_-, \tilde{p}_i)$ and $F_{r'}^{\mu(\nu)}(\tilde{p}_f, \tilde{q}_-)$ in the ratio (8) have the form:

$$F_n^{\mu'}(\tilde{p}_2, \tilde{p}_1) = a^{\mu'} L_n(\tilde{p}_2, \tilde{p}_1) + b_-^{\mu'} L_{n-1} + b_+^{\mu'} L_{n+1}, \tag{13}$$

$$n = r_1, r_2; \quad \mu' = \mu, \nu; \quad \tilde{p}_1 = \tilde{p}_i, \tilde{q}_-; \quad \tilde{p}_2 = \tilde{q}_-, \tilde{p}_f. \tag{14}$$

In this expression, the matrices $a^{\mu'}, b_\pm^{\mu'}$ are defined by the relations

$$a^{\mu'} = \tilde{\gamma}^{\mu'} + \eta^2 \frac{m^2}{2(k\tilde{p}_1)(k\tilde{p}_2)} k^{\mu'}\hat{k}, \tag{15}$$

$$b_\pm^{\mu'} = \frac{1}{4}\eta m \cdot \left[\frac{\hat{\varepsilon}_\pm \hat{k}\gamma^{\mu'}}{(k\tilde{p}_2)} + \frac{\gamma^{\mu'}\hat{k}\hat{\varepsilon}_\pm}{(k\tilde{p}_1)}\right], \quad \hat{\varepsilon}_\pm = \hat{e}_x \pm i\delta \cdot \hat{e}_y, \tag{16}$$

The special functions $L_n$ and $L_{n\pm 1}$, and their arguments have the form [21]:

$$L_n(\tilde{p}_2, \tilde{p}_1) = \exp(-in\chi_{\tilde{p}_2\tilde{p}_1}) \cdot J_n(\gamma_{\tilde{p}_2\tilde{p}_1}), \tag{17}$$

$$tg\,\chi_{\tilde{p}_2\tilde{p}_1} = \delta \cdot \frac{(e_y Q_{\tilde{p}_2\tilde{p}_1})}{(e_x Q_{\tilde{p}_2\tilde{p}_1})}, \quad Q_{\tilde{p}_2\tilde{p}_1} = \frac{\tilde{p}_2}{(k\tilde{p}_2)} - \frac{\tilde{p}_1}{(k\tilde{p}_1)}, \tag{18}$$

$$\gamma_{\tilde{p}_2\tilde{p}_1} = \eta m\sqrt{-Q_{\tilde{p}_2\tilde{p}_1}^2}, \tag{19}$$

In the future, we will consider the case when the initial and final particles have ultrarelativistic energies and fly in a narrow cone. In this case, the direction of wave propagation should be far from the specified narrow cone of particles (otherwise the resonances disappear [20, 24, 54-56]). Thus, the energies of electrons and gamma quanta must satisfy the conditions

$$E_j \gg m, \quad \omega_j \gg m, \quad j = i, f. \tag{20}$$

$$\theta_i = \angle(\mathbf{k}_i, \mathbf{p}_i) \ll 1, \quad \theta_f = \angle(\mathbf{k}_f, \mathbf{p}_f) \ll 1 \tag{21}$$

$$\theta = \angle(\mathbf{k}, \mathbf{k}_i) \sim 1, \quad \theta_{\mathbf{kp}_i} = \angle(\mathbf{k}, \mathbf{p}_i) \approx \theta \tag{22}$$

At the same time, we will assume that the value of the classical parameter $\eta$ (1) is limited from above by the value

$$\eta \ll \eta_{max}, \quad \eta_{max} = \frac{E_f}{m} \gg 1 \tag{23}$$

Therefore, further consideration of resonant processes will be valid for sufficiently high wave intensities. However, the intensity of these fields should be less than the critical Schwinger field $F_* \approx 1.3 \cdot 10^{16}$ V/cm [82, 87]. In this article, within the framework of condition (20), we will consider sufficiently large energies of initial electrons and not very large energies of initial gamma quanta:

$$E_i \gtrsim 1\,\text{GeV}, \quad \omega_i \ll 1\,\text{GeV} \tag{24}$$

At the same time, we will assume that the energy of the initial electron is much greater than the characteristic energy of the Compton effect, and the energy of the initial gamma quantum is less than or on the order of this energy:

$$E_i \gg \omega_C, \quad \omega_i \lesssim \omega_C. \tag{25}$$

Within the framework of conditions (24), (25), we determine the characteristic energy of the Compton effect (3) for various frequencies and intensities if an external electromagnetic wave propagates towards the momentum of the initial particles $(\theta = \pi)$.

$$\omega_C \approx \begin{cases} 130.56 \text{ MeV, if } \omega = 1 \text{ keV}, I = 1.86 \cdot 10^{24} \text{ Wcm}^{-2} (\eta = 1) \\ 26.112 \text{ MeV, if } \omega = 5 \text{ keV}, I = 4.65 \cdot 10^{25} \text{ Wcm}^{-2} (\eta = 1) \\ 6.528 \text{ MeV, if } \omega = 20 \text{ keV}, I = 7.44 \cdot 10^{26} \text{ Wcm}^{-2} (\eta = 1) \end{cases} \quad (26)$$

The resonant behavior of the amplitude (6)-(8) is due to the quasi-discrete structure of the system: an electron + a plane electromagnetic wave. Under resonance conditions, the intermediate electron enters the mass shell (Oleinik resonances) [33, 45-60]. Note that in this case, in addition to channels A and B, channel D is also possible, for which the intermediate particle is a positron (see Fig. 2). As a result, for intermediate electrons (channels A and B) or positrons (channel D), the laws of conservation of energy-momentum are fulfilled:

$$\tilde{q}_-^2 = m_*^2, \quad (27)$$

$$\tilde{q}_+^2 = m_*^2. \quad (28)$$

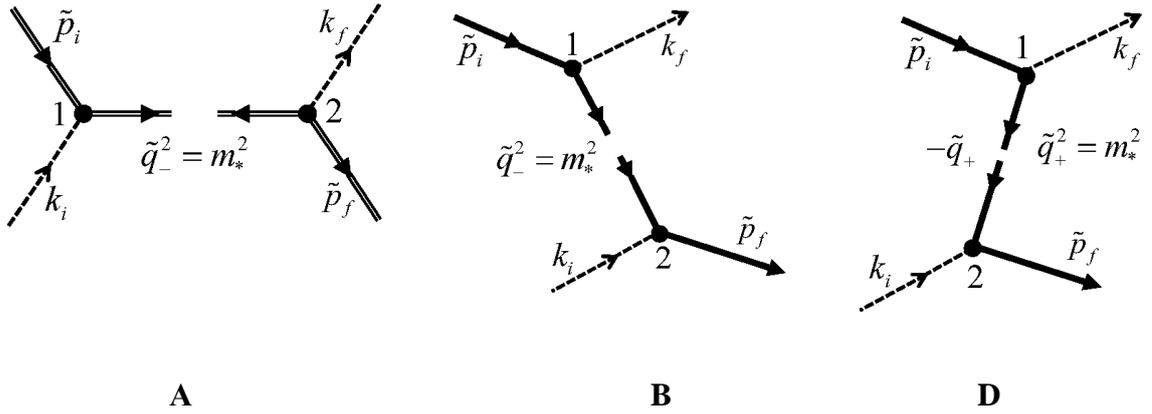

A            B            D

Fig. 2. Feynman diagrams of the resonant Compton effect in the field of a plane electromagnetic wave.

At the same time, the elimination of resonant infinity in the field of a flat monochromatic electromagnetic wave is carried out by the Breit-Wigner procedure [84, 20, 54-56]:

$$\left(\tilde{q}_\mp^2 - m_*^2\right)^2 \to \left(\tilde{q}_\mp^2 - m_*^2\right)^2 + 4m_*^2 \Gamma^2(\eta, \varepsilon) \quad (29)$$

In this case, the resonant width $\Gamma(\eta, \varepsilon)$ is determined by the expression

$$\Gamma(\eta, \varepsilon) = \frac{\tilde{q}_\mp^0}{2m_*} W(\eta, \varepsilon), \quad (30)$$

where $W(\eta, \varepsilon)$ is the total probability (per unit time) of the Compton effect stimulated by an external field on an intermediate electron or positron with a 4-momentum $\tilde{q}_\mp$ [19].

$$W(\eta, \varepsilon) = \frac{\alpha m^2}{4\pi \tilde{q}_\mp^0} \mathrm{K}(\eta, \varepsilon), \quad (31)$$

where

$$\mathrm{K}(\eta, \varepsilon) = \sum_{n=1}^\infty \int_0^{n\varepsilon} \frac{du}{(1+u)^2} K_n(\eta, u, n\varepsilon), \quad \varepsilon > 0. \quad (32)$$

Here the functions $K_n(u, n\varepsilon)$ are defined by the expression:

$$K_n(\eta, u, n\varepsilon) = -4J_n^2(\gamma) + \eta^2 \left(2 + \frac{u^2}{1+u}\right)\left(J_{n+1}^2 + J_{n-1}^2 - 2J_n^2\right), \quad (33)$$

$$\gamma = 2n \frac{\eta}{\sqrt{1+\eta^2}} \sqrt{\frac{u}{n\varepsilon}\left(1 - \frac{u}{n\varepsilon}\right)}. \tag{34}$$

Note that in expressions (32)-(34), the parameter $\varepsilon$ is determined by the corresponding reaction channel (see sections 3, 4, 5). Taking into account the ratios (30)-(32), the resonant width (30) will take the form:

$$\Gamma(\eta, \varepsilon) = \frac{\alpha m^2}{8\pi m_*} K(\eta, \varepsilon). \tag{35}$$

As will be shown below, under resonant conditions (27), (28) channels A, B and D may not interfere. Therefore, in the future we will consider sequentially the resonances of the Compton effect in a strong electromagnetic field for channels A, B and D in the absence of their interference. In section 5, it will be shown that in the energy range of the initial particles (24), the resonant channel A will be suppressed. Therefore, let's start studying the resonant Compton effect from channel B.

### 3. Resonant Compton effect in a strong field: channel B

For channel B, taking into account the resonant condition (27), the laws of conservation of the 4-momentum at the first and second vertices can be represented as follows (see Channel B in Fig.2):

$$\tilde{p}_i + r_1 k = \tilde{q}_- + k_f, \quad r_1 \geq 1 \tag{36}$$

$$\tilde{q}_- + k_i = \tilde{p}_f + r_2 k, \quad r_2 \geq 1 \tag{37}$$

Hence, and from the type of amplitude (6)-(8), it follows that for channel B under resonant conditions, the second-order process according to the fine structure constant is effectively reduced to two first-order processes of the type of Compton effect stimulated by an external electromagnetic field. At the first vertex, the absorption of the $r_1$- wave photons by the initial electron and the emission of the intermediate electron and the final gamma quantum take place. At the second vertex, we obtain the scattering of the initial gamma quantum on an intermediate electron with radiation $r_2$- photons of the wave and the final electron.

The expression for the resonant frequency $\left(\omega_{B(r_1)}\right)$ of the scattered gamma quantum in the case of channel B (see Fig.2B) is obtained taking into account the conservation law of the 4-momentum (36) for the first vertex, as well as conditions (20), (22) и (27):

$$\omega_{B(r_1)} = E_i \frac{r_1 \varepsilon_{iC}}{r_1 \varepsilon_{iC} + \left(1 + \delta_{fi}^2\right)} < E_i. \tag{38}$$

Here, the quantum parameter $\varepsilon_{iC}$ is determined by the expression (2), (3), and the ultrarelativistic parameter $\delta_{fi}^2$, which determines the outgoing angle of the final gamma quantum relative to the momentum of the initial electron, is equal to

$$\delta_{fi}^2 = \frac{E_i^2 \theta_{fi}^2}{m_*^2}, \quad \theta_{fi} = \angle(\mathbf{k}_f, \mathbf{p}_i) \ll 1. \tag{39}$$

It is important to note that for channel B, the resonant frequency of the final gamma quantum is determined by the outgoing angle ($\delta_{fi}^2$), the number of absorbed photons of the wave $(r_1)$, as well as the quantum parameter $\varepsilon_{iC}$. At the same time, the energy of the final gamma quantum for channel B is always less than the initial energy of the initial electron. As will be shown below, the opposite situation will be for channel D.

Carrying out similar calculations for the second vertex, from the ratios (27), (37) we obtain an expression for the resonant energy of the final electron $\left(E_{B(r_2)}\right)$ in the case of channel B:

$$E_{B(r_2)} = \omega_i \frac{r_2\varepsilon'_{iC} + \sqrt{(r_2\varepsilon'_{iC})^2 + 4(r_2\varepsilon'_{iC} - \delta'^2_{fi})}}{2(r_2\varepsilon'_{iC} - \delta'^2_{fi})} > \omega_i, \quad (40)$$

Here, the quantum parameter $\varepsilon'_{iC}$ is determined by the expression (2), (3), and the ultrarelativistic parameter $\delta'^2_{fi}$, which determines the outgoing angle of the final electron relative to the momentum of the initial gamma quantum, is equal to

$$\delta'^2_{fi} = \frac{\omega_i^2 \theta'^2_{fi}}{m_*^2}, \quad \theta'_{fi} = \angle(\mathbf{p}_f, \mathbf{k}_i) \ll 1. \quad (41)$$

It can be seen from expression (40) that the resonant energy of a final electron depends on its outgoing angle $\left(\delta'^2_{fi}\right)$, the number of emitted photons of the wave $(r_2)$ and the quantum parameter $\varepsilon'_{iC}$. It is important to note that for channel B, the resonant energy of the final electron always exceeds the energy of the initial gamma quantum. As will be shown below, for channel D, the energy of the final electron will be less than the initial energy of the gamma quantum. It should be borne in mind that in expression (40), the ultrarelativistic parameter $\delta'^2_{fi}$ should not come very close to the value of the parameter $r_2\varepsilon'_{iC}$. This is due to the fact that the law of conservation of energy must be fulfilled for channel B

$$E_{B(r_2)} + \omega_{B(r_1)} \approx E_0 \quad (42)$$

Substituting the energy of the electron (40) into expression (42), after simple transformations, we obtain a connection of ultrarelativistic parameters that determine the outgoing angles of the final electron and the gamma quantum:

$$\delta'^2_{fi} = r_2\varepsilon'_{iC}\left\{1 - \frac{\varepsilon'_{iC}\beta_{fi(r_1)}}{(1+\varepsilon'_{iC}\beta_{fi(r_1)})}\left[1 + \frac{\beta_{fi(r_1)}}{2r_2(1+\varepsilon'_{iC}\beta_{fi(r_1)})}\right]\right\}. \quad (43)$$

It is indicated here:

$$\beta_{fi(r_1)} = \frac{1}{\varepsilon_{iC}} + \frac{r_1}{1+\delta^2_{fi}}. \quad (44)$$

At the same time, the expression in curly brackets in the ratio (43) will be positive if the condition for the number of emitted photons of the wave at the second vertex is met (see ratio (63)). We emphasize that the ratio (43) gives an unambiguous relationship between the outgoing angles of the final electron and the gamma quantum. Figure 3 shows the dependence of the square of the outgoing angle of a final electron on the square of the outgoing angle of a final gamma quantum in a strong X-ray wave field for a different number of absorbed (at the first vertex) and emitted (at the second vertex) photon waves. This connection is a distinctive feature of the resonant process, in contrast to the non-resonant process, where there is no such connection of the outgoing angles of the final particles.

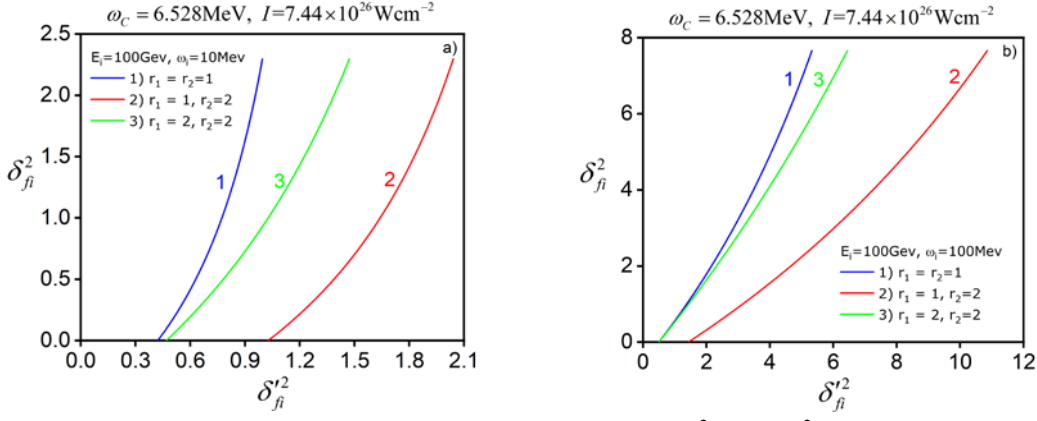

Fig. 3. Dependence of the ultrarelativistic parameters $\delta'^2_{fi}$ and $\delta^2_{fi}$ (43), (44) determining the squares of the outgoing angles of the electron and gamma quantum for a different number of emitted and absorbed photons of the X-ray wave at fixed values of the initial energy of the electron and the characteristic energy of the Compton effect and two possible values of the initial energy of the gamma quantum.

It is important to note that when the energy of the initial electron significantly exceeds the characteristic energy of the Compton effect (see relations (25)), then the resonant energy of the final gamma quantum (38) tends to the energy of the initial electron from the bottom:

$$\omega_{B(r_1)} \approx E_i\left(1 - \frac{1+\delta^2_{fi}}{\varepsilon_{iC(r_1)}}\right) = E_i - \frac{1}{r_1}\left(1+\delta^2_{fi}\right)\omega_C \approx E_i \quad (E_i \gg \omega_C). \tag{45}$$

At the same time, the resonant energy of the final electron, according to the ratios (42), (45), can be determined by the outgoing angle of the final gamma quantum and the number of absorbed photons of the wave at the first vertex $r_1$:

$$E_{B(r_1)} \approx \omega_i + \frac{1}{r_1}\left(1+\delta^2_{fi}\right)\omega_C \tag{46}$$

In the future, we will study the resonant cross section in the absence of interference of channels A, B and D. As will be shown below, this is possible, since in the energy region (24) channel A will be suppressed. At the same time, for resonant channels, the energies of final particles lie in different energy regions. In addition, within each channel (B or D), processes with different numbers of absorbed and emitted photons of the wave also have different energies and do not interfere. Because of this, in the future we will consider the resonant cross section separately for each of the reaction channels.

Using an expression for the amplitude of the process (see expressions (5)-(8), (13)), It is not difficult to obtain a resonant differential cross section in the case of unpolarized particles and the absence of interference of reaction channels. After the standard calculations [85] for channel B, we get:

$$d\sigma_{B(r_1 r_2)} = r_e^2 \frac{2m^6 E_i}{(4\pi)^2 \omega_i \omega_f E_f m_*^2 \left(1+\delta_i^2\right)} \frac{K_{r_1}(u_{1B}, v_{1B}) K_{r_2}(u_{2B}, v_{2B})}{\left[\left(\tilde{q}_-^2 - m_*^2\right)^2 + 4m_*^2 \Gamma^2(\eta, \varepsilon_{iC})\right]} \times$$
$$\times \delta^{(4)}\left[\tilde{p}_f + k_f - \tilde{p}_i - k_i - (r_2 - r_1)k\right] d^3\tilde{p}_f d^3 k_f. \tag{47}$$

When obtaining a resonant differential cross section, the corresponding probability (per unit of time and unit of volume) was divided by the flux density of the initial particles [85]:

$$j_i = \frac{(k_i \tilde{p}_i)}{\omega_i \tilde{E}_i} \approx \frac{m_*^2}{2E_i^2}\left(1+\delta_i^2\right), \quad \delta_i^2 = \frac{E_i^2 \theta_i^2}{m_*^2}. \tag{48}$$

Here $\theta_i$ is the angle of the solution between the momenta of the initial particles (21). In expression (47), $\Gamma(\eta, \varepsilon_{iC})$ this is the resonant width, which is determined by expression (35). The functions $K_{r_1}(u_{1B}, v_{1B})$ and $K_{r_2}(u_{2B}, v_{2B})$ determine the probability of the Compton effect stimulated by an external field at the first and second vertices (see Fig. 2B)

$$K_{r_j}(u_{jB}, v_{jB}) = -4J_{r_j}^2(\gamma_{jB}) + \eta^2 \left(2 + \frac{u_{jB}^2}{1 + u_{jB}}\right)\left(J_{r_j+1}^2 + J_{r_j-1}^2 - 2J_{r_j}^2\right), \quad j = 1, 2. \quad (49)$$

At the same time, the arguments of the Bessel functions and the corresponding relativistically invariant parameters $u_{jB}, v_{jB}$ are defined by expressions:

$$\gamma_{jB} = 2r_j \frac{\eta}{\sqrt{1+\eta^2}} \sqrt{\frac{u_{jB}}{v_{jB}}\left(1 - \frac{u_{jB}}{v_{jB}}\right)}, \quad j = 1, 2, \quad (50)$$

$$u_{1B} = \frac{(kk_f)}{(kq_-)} \approx \frac{\omega_f}{(E_i - \omega_f)}, \quad v_{1B} = 2r_1 \frac{(kp_i)}{m_*^2} \approx r_1 \varepsilon_{iC}, \quad (51)$$

$$u_{2B} = \frac{(kk_i)}{(kq_-)} \approx \frac{\omega_i}{E_f - \omega_i} = \frac{\omega_i}{E_i - \omega_f}, \quad (52)$$

$$v_{2B} = 2r_2 \frac{(kp_f)}{m_*^2} \approx r_2 \varepsilon'_{iC} \frac{E_f}{\omega_i} = r_2 \varepsilon'_{iC} \frac{(E_0 - \omega_f)}{\omega_i}. \quad (53)$$

The four-dimensional Dirac delta function in expression (47) makes it easy to integrate the resonant cross section in terms of energy and azimuthal angle of a final gamma quantum. Taking this into account, the resonant differential cross section under conditions (20)-(23) will take the following form:

$$R_{B(r_1 r_2)} = \frac{d\sigma_{(r_1 r_2)}}{d\delta_{fi}^2} = \frac{r_e^2}{8\pi(1+\delta_i^2)}\left(\frac{m^4}{m_*^4}\right) \frac{m^2 E_i}{\omega_i \omega_f (E_0 - \omega_f)} \frac{K_{r_1}(u_{1B}, v_{1B}) K_{r_2}(u_{2B}, v_{2B})}{\left[\left(\delta_{fi(0)}^2 - \delta_{fi}^2\right)^2 + \Upsilon_{fi}^2\right]}. \quad (54)$$

Here $\delta_{fi(0)}^2$ is an ultrarelativistic parameter that varies independently of the resonant frequency of the gamma quantum, and the ultrarelativistic parameter $\delta_{fi}^2$ (39) is related to the resonant frequency by the ratio (38). The value $\Upsilon_{fi}$ is the angular resonant width for the channel B.

$$\Upsilon_{fi} = \frac{\alpha m^2}{4\pi m_*^2}\left(\frac{E_i}{\omega_f}\right) K(\eta, \varepsilon_{iC}) \quad (55)$$

When the condition is met

$$\left(\delta_{fi(0)}^2 - \delta_{fi}^2\right)^2 \ll \Upsilon_{fi}^2, \quad (56)$$

then in expression (54) we can put $\omega_f = \omega_{B(r_1)}$. As a result, after simple transformations, we obtain the maximum resonant differential cross section for the channel B:

$$R_{B(r_1 r_2)}^{\max} = \frac{d\sigma_{B(rr')}^{\max}}{d\delta_{fi}^2} = r_e^2 c_B \Psi_{B(r_1 r_2)}. \quad (57)$$

Here the function $c_B$ is determined by the initial installation parameters

$$c_B = \frac{2\pi}{\alpha^2 (1+\delta_i^2) \varepsilon'_{iC} K^2(\eta, \varepsilon_{iC})} \left(\frac{m}{\omega_C}\right)^2 \approx$$
$$\approx \frac{1.18 \cdot 10^5}{(1+\delta_i^2) \varepsilon'_{iC} K^2(\eta, \varepsilon_{iC})} \left(\frac{m}{\omega_C}\right)^2, \quad (58)$$

and the functions $\Psi_{B(rr')}$ determine the spectral-angular distribution of the resonant scattering cross section for the channel B:

$$\Psi_{B(r_1 r_2)} = \frac{r_1 \cdot K_{r_1}(u_{1B}, v_{1B}) K_{r_2}(u_{2B}, v_{2B})}{r_1 \varepsilon'_{iC} + (1+\delta_{fi}^2)(E_0/E_i)}. \quad (59)$$

Here, the relativistically invariant parameters (51), (50) for the first vertex of the Feynman diagram (Figure 2B) take the following form:

$$u_{1B} \approx \frac{r_1 \varepsilon_{iC}}{(1+\delta_{fi}^2)}, \quad v_{1B} \approx r_1 \varepsilon_{iC}, \quad (60)$$

$$\gamma_{1B} = 2 r_1 \frac{\eta}{\sqrt{1+\eta^2}} \frac{\delta_{fi}}{(1+\delta_{fi}^2)}. \quad (61)$$

For the second vertex, the relativistically invariant parameters (52), (53) take the form:

$$u_{2B} \approx \varepsilon'_{iC} \beta_{fi(r_1)} \quad v_{2B} \approx r_2 \left( \varepsilon'_{iC} + \frac{1}{\beta_{fi(r_1)}} \right). \quad (62)$$

Here the function $\beta_{fi(r_1)}$ is defined by the expression (44). At the same time, due to the coordination of processes in the first and second vertices, it is necessary to require the fulfillment of a condition $u_{2B} \leq v_{2B}$ that guarantees the validity of Bessel's arguments $\gamma_{2B}$ (50) in functions $K_{r_2}(u_{2B}, v_{2B})$. This process matching condition can be written as a condition for the allowed number of emitted photons of the wave at the second vertex:

$$r_2 \geq r_{B(r_1)}^{min} = \lceil r_{B(r_1)} \rceil, \quad r_{B(r_1)} = \frac{\beta_{max(r_1)}}{1+(\varepsilon'_{iC} \beta_{max(r_1)})^{-1}}, \quad \beta_{max(r_1)} = r_1 + \frac{1}{\varepsilon_{iC}}. \quad (63)$$

Note that the condition (63) for the number of photons of the wave at the second vertex ensures the positivity of the ultrarelativistic parameter $\delta_{fi}^2$ in the ratio (43). Thus, conditions (43) and (63) uniquely determine the dependence of the outgoing angles of the final electron and the gamma quantum in the resonant case (see Figure 3). It is important to emphasize that for channel B, the number of absorbed photons of the wave at the first vertex can be arbitrary $(r_1 \geq 1)$, and the number of emitted photons at the second vertex is limited from below by the condition (63).

Consider the case when the energy of the initial electron significantly exceeds the characteristic energy of the Compton effect (see ratio (25)). Then, the quantum parameter $\varepsilon_{iC} \gg 1$ and even with a small number of absorbed photons of the wave $r_1 \geq 1$, the resonant frequency of the final gamma quantum will be close to the energy of the initial electron (see expression (45)). At the same time, the parameter $u_{1B}$ (60) will be much larger than one. As a result, the expression for functions $K_{r_1}(u_{1B}, v_{1B})$ is simplified

$$K_{r_1}(u_{1B}, v_{1B}) \approx \varepsilon_{iC} \eta^2 \frac{r_1}{(1+\delta_{fi}^2)} \left[ J_{r_1+1}^2 + J_{r_1-1}^2 - 2 J_{r_1}^2(\gamma_{1B}) \right] \quad (64)$$

Now let's simplify the expressions for relativistically invariant parameters (62). After simple calculations, we get:

$$u_{2B} \approx \frac{r_1 \varepsilon'_{iC}}{1+\delta^2_{fi}}, \quad v_{2B} \approx r_2 \left[\varepsilon'_{iC} + \frac{(1+\delta^2_{fi})}{r_1}\right] \tag{65}$$

Considering this, after simple transformations, the expression for the maximum resonant section (57) will take the form:

$$R'^{\max}_{B(r_1 r_2)} == r_e^2 c'_B \Psi'_{B(r_1 r_2)} \quad (E_i \gg \omega_C) \tag{66}$$

where the functions $c'_B$ and $\Psi'_{B(r_1 r_2)}$ have the following form:

$$c'_B = \frac{2\pi\eta^2}{\alpha^2(1+\delta_i^2)}\left(\frac{m}{\omega_C}\right)^2 \frac{\varepsilon_{iC}}{\varepsilon'_{iC} K^2(\eta, \varepsilon_{iC})}, \tag{67}$$

$$\Psi'_{B(r_1 r_2)} = \frac{r_1^2 \left[J^2_{r_1+1} + J^2_{r_1-1} - 2J^2_{r_1}(\gamma_{1B})\right] K_{r_2}(u_{2B}, v_{2B})}{(1+\delta^2_{fi})\left[(1+\delta^2_{fi})(E_0/E_i) + r_1\varepsilon'_{iC}\right]} \tag{68}$$

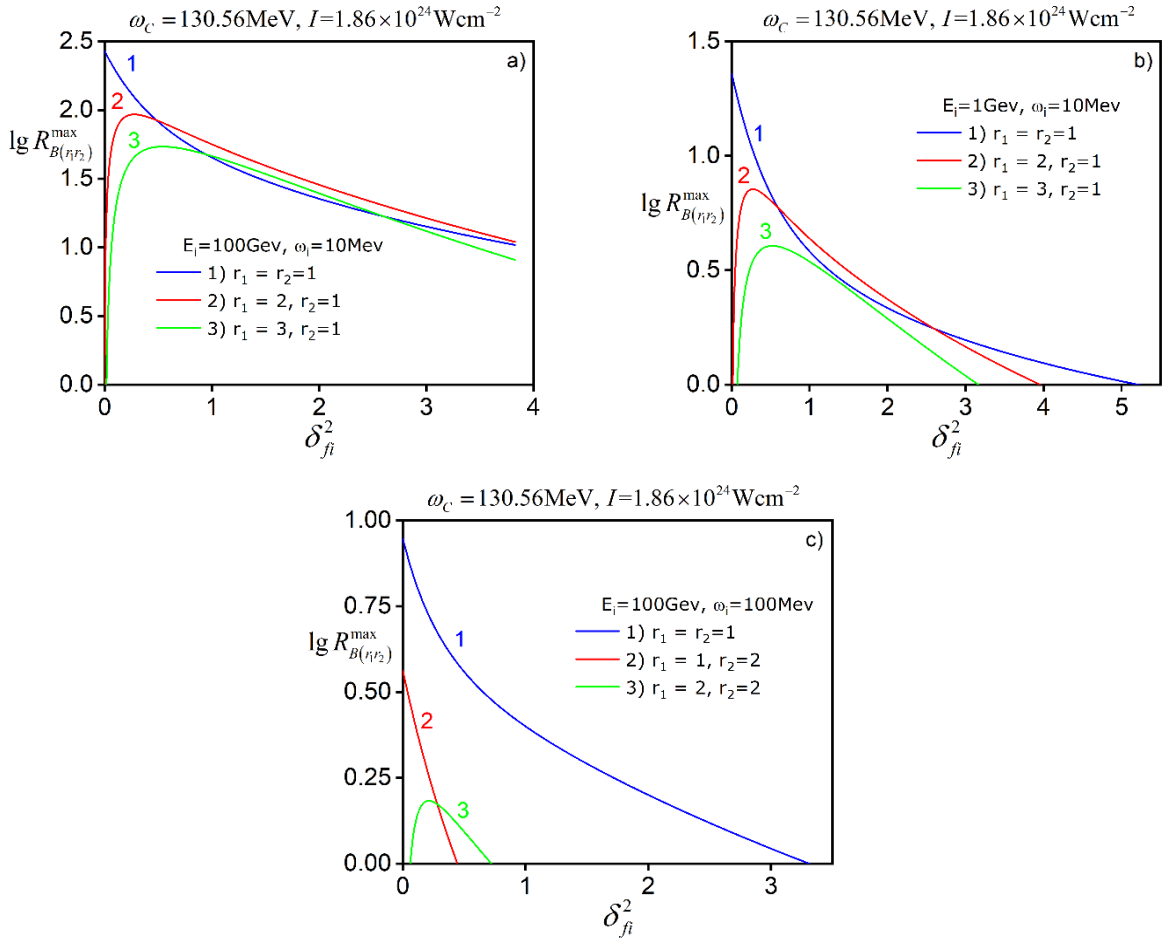

Fig. 4. Dependence of the maximum resonant cross section (57) on the ultrarelativistic parameter $\delta^2_{fi}$ for various energies of the initial particles, as well as the numbers of emitted and absorbed photons of the X-ray wave. Case a) corresponds to the values of the parameters $\varepsilon_{iC} \approx 7.66 \times 10^2, \varepsilon'_{iC} \approx 0.08$; case b) - $\varepsilon_{iC} \approx 7.66, \varepsilon'_{iC} \approx 0.08$; case c) - $\varepsilon_{iC} \approx 7.66 \times 10^2, \varepsilon'_{iC} \approx 0.8$.

Here, the parameter $\gamma_{1B}$ is given by the expression (61), and the parameters $u_{2B}, v_{2B}$ have the form (65). It is important to note that the maximum order of magnitude of the resonant section (66) (at $r_1 = r_2 = 1$) is determined by the function $c'_B$ (see Tables 1-3). Therefore, you can write

$$R'^{\max}_{B(r_1 r_2)} \lesssim c'_B r_e^2. \tag{69}$$

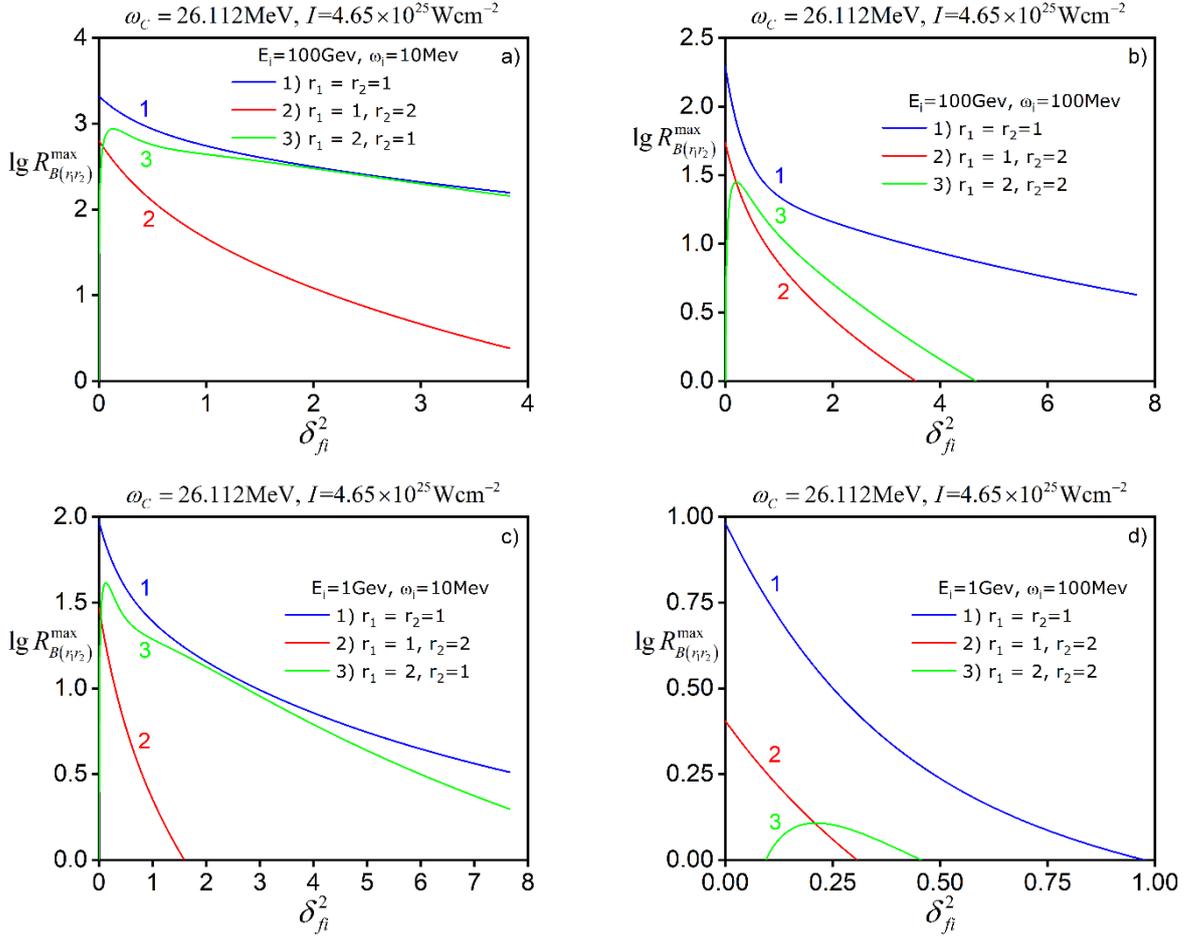

Fig. 5. Dependence of the maximum resonant cross section (57) on the ultrarelativistic parameter $\delta^2_{fi}$ for various energies of the initial particles, as well as the numbers of emitted and absorbed photons of the X-ray wave.

Figures 4-6 show the dependences of the resonance section (57)-(59) on the square of the outgoing angle of the final gamma quantum for various energies of the initial particles, as well as various characteristic energies of the Compton effect and the numbers of absorbed $(r_1)$ and emitted $(r_2)$ photons of the wave. Tables 1, 2, 3 show the values of the maximum resonant cross section corresponding to the peaks in Figures 4-6, as well as the corresponding values of the resonant energies of final gamma quanta and electrons. It can be seen from these figures and tables that the resonant cross section has the largest value for the numbers of absorbed and emitted photons of the wave $r_1 = r_2 = 1$ at zero outgoing angle of the final gamma quantum relative to the momentum of the initial electron. With an increase in the number of absorbed and emitted photons of the wave, the resonant cross section decreases, and the maximum angular distribution of the resonant cross section shifts towards large angles. At the same time, the magnitude of the resonant cross section depends very much on the ratio of the energies of the initial particles to the characteristic energy of the Compton effect (parameters $\varepsilon_{iC}$ and $\varepsilon'_{iC}$). So, for $r_1 = r_2 = 1$ and the energies of the initial particles $E_i = 100$ GeV, $\omega_i = 10$ MeV, (see cases a) in Figures 4-6) at parameter values $(\varepsilon_{iC}, \varepsilon'_{iC}) \approx (7.66 \times 10^2, 0.08)$, $(3.83 \times 10^3, 0.34)$, $(1.52 \times 10^4, 1.53)$, the maximum resonant differential cross sections and the corresponding gamma quantum energies take the following

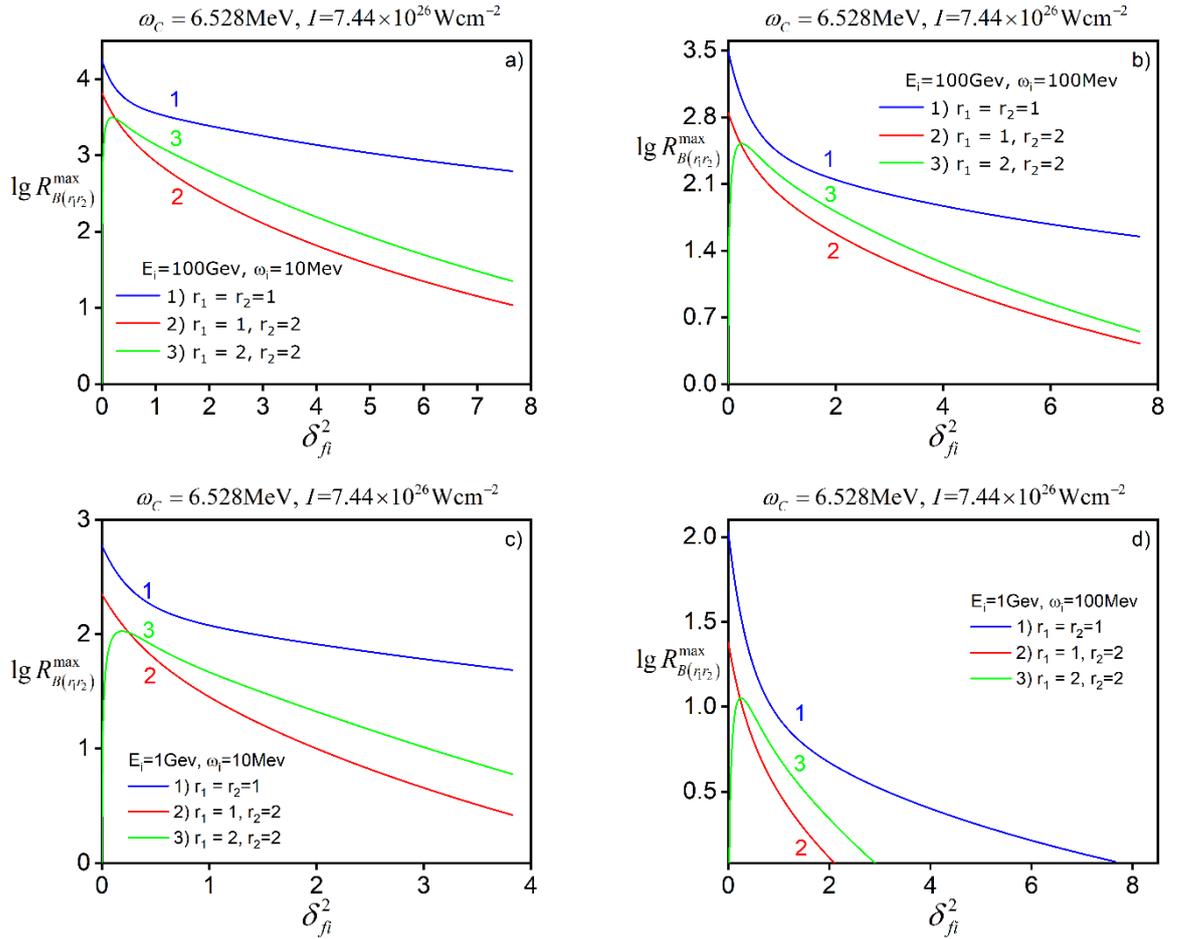

Fig. 6 Dependence of the maximum resonant cross section (57) on the ultrarelativistic parameter $\delta_{fi}^2$ for various energies of the initial particles, as well as the numbers of emitted and absorbed photons of the X-ray wave.

Table 1.

$\omega_C = 130.56$ MeV, $\omega = 1$ keV, $I = 1.86 \times 10^{24}$ Wcm$^{-2}$

| (GeV) | $r_1, r_2$ | $\delta_{fi}^{2(*)}$ | $\omega_B$ (GeV) | $\delta'^{2(*)}_{fi}$ | $E_B$ (GeV) | $R^{\max(*)}_{B(r_1 r_2)}$ $(r_e^2)$ | $c'_B$ |
|---|---|---|---|---|---|---|---|
| $E_i = 100$ $\omega_i = 0.01$ | 1, 1 | 0 | 99.87 | 0.069 | 0.14 | $2.68 \times 10^2$ | $1.78 \times 10^2$ |
| | 2, 1 | 0.28 | 99.92 | 0.063 | 0.09 | $0.93 \times 10^2$ | |
| | 3, 1 | 0.54 | 99.93 | 0.058 | 0.08 | 54.22 | |
| $E_i = 1$ $\omega_i = 0.01$ | 1, 1 | 0 | 0.88 | 0.067 | 0.13 | 22.91 | 17.8 |
| | 2, 1 | 0.27 | 0.92 | 0.061 | 0.09 | 7.15 | |
| | 3, 1 | 0.52 | 0.94 | 0.056 | 0.07 | 4.05 | |
| $E_i = 100$ $\omega_i = 0.1$ | 1, 1 | 0 | 99.87 | 0.339 | 0.23 | 8.86 | 14.17 |
| | 1, 2 | 0 | 99.87 | 0.773 | 0.23 | 3.66 | |
| | 2, 2 | 0.21 | 99.92 | 0.520 | 0.18 | 1.52 | |

values, respectively: $R_{B(1,1)}^{\max} \approx 2.68 \times 10^2,\ 2.09 \times 10^3,\ 1.76 \times 10^4\ (r_e^2)$ и $\omega_{B(1)} \approx 99.87,\ 99.97,\ 99.99\ (\text{GeV})$. In this case, the function $c_B'$ (67) takes the following values, respectively: $c_B' = 1.78 \times 10^2,\ 2.96 \times 10^3,\ 3.53 \times 10^4$. Thus, the resonant differential cross section increases with increasing parameter $\varepsilon_{iC}$, and in order of magnitude in accordance with the change in function $c_B'$.

Table 2.
$\omega_C = 26.112$ MeV, $\omega = 5$ keV, $I = 4.65 \cdot 10^{25}$ Wcm$^{-2}$

| (GeV) | $r_1, r_2$ | $\delta_{fi}^{2(*)}$ | $\omega_B$ (GeV) | $\delta_{fi}'^{2(*)}$ | $E_B$ (GeV) | $R_{B(r_1 r_2)}^{\max(*)}$ $(r_e^2)$ | $c_B'$ |
|---|---|---|---|---|---|---|---|
| $E_i = 100$ $\omega_i = 0.01$ | 1, 1 | 0 | 99.974 | 0.24 | 0.036 | $2.09 \times 10^3$ | $2.96 \times 10^3$ |
| | 2, 1 | 0.13 | 99.985 | 0.15 | 0.025 | $0.87 \times 10^3$ | |
| | 1, 2 | 0 | 99.974 | 0.52 | 0.036 | $6.32 \times 10^2$ | |
| $E_i = 100$ $\omega_i = 0.1$ | 1, 1 | 0 | 99.974 | 0.48 | 0.126 | $1.97 \times 10^2$ | $2.96 \times 10^2$ |
| | 1, 2 | 0 | 99.974 | 1.27 | 0.126 | 54.82 | |
| | 2, 2 | 0.21 | 99.984 | 0.67 | 0.116 | 28.11 | |
| $E_i = 1$ $\omega_i = 0.01$ | 1, 1 | 0 | 0.975 | 0.24 | 0.035 | 94.10 | 134.01 |
| | 2, 1 | 0.13 | 0.985 | 0.14 | 0.025 | 41.22 | |
| | 1, 2 | 0 | 0.975 | 0.51 | 0.035 | 29.65 | |
| $E_i = 1$ $\omega_i = 0.1$ | 1, 1 | 0 | 0.974 | 0.46 | 0.126 | 9.58 | 13.40 |
| | 1, 2 | 0 | 0.974 | 1.24 | 0.126 | 2.54 | |
| | 2, 2 | 0.21 | 0.984 | 0.66 | 0.116 | 1.28 | |

Table 3.
$\omega_C = 6.528$ MeV, $\omega = 20$ keV, $I = 7.44 \cdot 10^{26}$ Wcm$^{-2}$

| (GeV) | $r_1, r_2$ | $\delta_{fi}^{2(*)}$ | $\omega_B$ (GeV) | $\delta_{fi}'^{2(*)}$ | $E_B$ (GeV) | $R_{B(r_1 r_2)}^{\max(*)}$ $(r_e^2)$ | $c_B'$ |
|---|---|---|---|---|---|---|---|
| $E_i = 100$ $\omega_i = 0.01$ | 1, 1 | 0 | 99.993 | 0.422 | 0.016528 | $1.76 \times 10^4$ | $3.53 \times 10^4$ |
| | 1, 2 | 0 | 99.993 | 1.027 | 0.016528 | $0.66 \times 10^4$ | |
| | 2, 2 | 0.19 | 99.996 | 0.5971 | 0.013881 | $3.14 \times 10^3$ | |
| $E_i = 100$ $\omega_i = 0.1$ | 1, 1 | 0 | 99.993 | 0.4981 | 0.107 | $3.13 \times 10^3$ | $3.53 \times 10^3$ |
| | 1, 2 | 0 | 99.993 | 1.4367 | 0.107 | $0.71 \times 10^3$ | |
| | 2, 2 | 0.25 | 99.996 | 0.7413 | 0.104 | $3.33 \times 10^2$ | |
| $E_i = 1$ $\omega_i = 0.01$ | 1, 1 | 0 | 0.99351 | 0.4187 | 0.016486 | $5.98 \times 10^2$ | $1.19 \times 10^3$ |
| | 1, 2 | 0 | 0.99351 | 1.0213 | 0.016486 | $2.24 \times 10^2$ | |
| | 2, 2 | 0.19 | 0.99614 | 0.5935 | 0.013863 | $1.06 \times 10^2$ | |
| $E_i = 1$ $\omega_i = 0.1$ | 1, 1 | 0 | 0.993 | 0.4921 | 0.107 | $1.08 \times 10^2$ | $1.19 \times 10^2$ |
| | 1, 2 | 0 | 0.993 | 1.4251 | 0.107 | 24.05 | |
| | 2, 2 | 0.25 | 0.996 | 0.7384 | 0.104 | 11.25 | |

## 4. Resonant Compton effect in a strong field: channel D

For channel D, the intermediate electron becomes a positron. Therefore, in the ratios (36) and (37), it is necessary to make a replacement:

$$\tilde{q}_- \to -\tilde{q}_+, \quad r_1 \to -r_1, \quad r_2 \to -r_2 \tag{70}$$

Taking this into account, as well as the resonant condition (28), we obtain the following laws of conservation of energy-momentum at the first and second vertices of the Feynman diagram (see channel D in Fig. 2):

$$\tilde{q}_+ + \tilde{p}_i = k_f + r_1 k, \quad r_1 \geq 1, \tag{71}$$

$$k_i + r_2 k = \tilde{p}_f + \tilde{q}_+, \quad r_2 \geq 1. \tag{72}$$

Thus, at the second vertex (72), the external field-stimulated Breit-Wheeler process takes place (the generation of an intermediate positron and a final electron by the initial gamma quantum and $r_2$-photons). At the same time, at the first vertex (71), the reverse external field-stimulated Breit-Wheeler process takes place (annihilation of the intermediate positron and the initial electron into the final gamma quantum and $r_1$-photons of the wave). Given the relations (28), (71), it is possible to obtain an expression for the energy of a final gamma quantum $\omega_{D(r_1)}$:

$$\omega_{D(r_1)} = \frac{E_i}{\xi_{fi(r_1)}} > E_i. \tag{73}$$

It is indicated here:

$$\xi_{fi(r_1)} = 1 - \frac{(1+\delta_{fi}^2)}{r_1 \varepsilon_{iC}} \quad \left(0 < \xi_{fi(r_1)} < 1\right) \tag{74}$$

Here the values $\delta_{fi}^2$ and $\varepsilon_{iC}$ are determined by expressions (39) and (2). It is important to emphasize that the resonant energy of the final gamma quantum for channel D exceeds the energy of the initial electron. In addition, unlike channel B (see expression (38)), expression (73) implies restrictions on the values of parameters $\delta_{fi}^2$ and $r_1 \varepsilon_{iC}$:

$$0 \leq \delta_{fi}^2 < \delta_{\max(r_1)}^2, \quad \delta_{\max(r_1)}^2 = (r_1 \varepsilon_{iC} - 1) > 0, \tag{75}$$

At the same time, the inequality must be fulfilled

$$r_1 \varepsilon_{iC} > 1. \tag{76}$$

Note that conditions (75) and (76) are necessary, but insufficient. As will be shown below, the coordination of resonant processes in the first and second vertices of channel D will lead to more stringent conditions for the interval of change in the ultrarelativistic parameter $\delta_{fi}^2$ and the number of emitted photons of the wave at the first vertex (see relations (89) and (88)).

Consider the case when the quantum parameter

$$r_1 \varepsilon_{iC} \gg 1, \tag{77}$$

then, for not very large values of the ultrarelativistic parameter $\delta_{fi}^2$ $\left(\delta_{fi}^2 \ll r_1 \varepsilon_{iC}\right)$, the expression for the energy of the final gamma quantum (73) tends to the energy of the initial electron from above (compare with the corresponding expression for the energy of the final gamma quantum for channel B (45)):

$$\omega_{D(r_1)} \approx E_i \left[1 + \frac{1+\delta_{fi}^2}{\varepsilon_{iC(r_1)}}\right] = E_i + \frac{1}{r_1}(1+\delta_{fi}^2)\omega_C \approx E_i \quad (E_i \gg \omega_C). \tag{78}$$

Using the relations (27) and (72), it is possible to obtain an expression for the resonant energy of an electron in the external field-stimulated Breit-Wheeler process (see also [55, 58]):

$$E_{D(r_2)} = \omega_i \frac{r_2\varepsilon'_{iBW} \pm \sqrt{r_2\varepsilon'_{iBW}(r_2\varepsilon'_{iBW} - 1) - \tilde{\delta}^2_{fi}}}{2(r_2\varepsilon'_{iBW} + \tilde{\delta}^2_{fi})} < \omega_i. \tag{79}$$

Here, the quantum parameter $\varepsilon'_{iBW}$ is determined by the expression (2), (3), and the ultrarelativistic parameter $\tilde{\delta}'^2_{fi}$, which determines the outgoing angle of the final electron relative to the momentum of the initial gamma quantum, is equal to

$$\tilde{\delta}^2_{fi} = \frac{1}{4}\delta'^2_{fi} = \frac{\omega_i^2 \theta'^2_{fi}}{4m_*^2}, \tag{80}$$

It can be seen from expression (79) that the resonant energy of a final electron depends on its outgoing angle $(\tilde{\delta}'^2_{fi})$, the number of absorbed photons of the wave $(r_2)$ and the quantum parameter $\varepsilon'_{iBW}$. Thus, for channel D, the energy of the final gamma quantum is determined by the characteristic energy of the Compton effect, and the energy of the electron is determined by the characteristic Breit-Wheeler energy.

It follows from expression (79) that the ultrarelativistic parameter $\tilde{\delta}^2_{fi}$ determining the square of the outgoing angle of the final electron relative to the momentum of the initial gamma quantum can vary in the interval:

$$0 \leq \tilde{\delta}^2_{fi} \leq \tilde{\delta}^2_{\max(r_2)}, \quad \tilde{\delta}^2_{\max(r_2)} = r_2\varepsilon'_{iBW}(r_2\varepsilon'_{iBW} - 1) \tag{81}$$

Hence the inequality follows:

$$r_2\varepsilon'_{iBW} \geq 1 \tag{82}$$

Due to condition (82), depending on the ratio between the energy of the initial gamma quantum and the characteristic Breit-Wheeler energy, the resonant process through channel D can take place with a different number of absorbed photons of the wave:

$$r_2 \geq r_{2D}^{\min} = \lceil \varepsilon'^{-1}_{iBW} \rceil > 1, \text{ if } \omega_i < \omega_{BW}, \tag{83}$$

$$r_2 \geq 1, \text{ if } \omega_i \geq \omega_{BW} \tag{84}$$

The expression for the energy of the final electron (79) for each value of the ultrarelativistic parameter $\tilde{\delta}^2_{fi}$ in the range (81) takes two possible values. At the same time, for $\tilde{\delta}^2_{fi} = 0$ the electron energy can take both maximum $E^+_{D(r_2)}$ and minimum $E^-_{D(r_2)}$ values:

$$E^{\pm}_{D(r_2)} = \frac{\omega_i}{2}\left[1 \pm \sqrt{1 - \frac{1}{r_2\varepsilon'_{iBW}}}\right], \tag{85}$$

Considering the ratios (73) and (79), as well as the law of conservation of energy

$$E_{D(r_2)} + \omega_{D(r_1)} \approx E_0, \tag{86}$$

it is possible to obtain a connection between the outgoing angles of an electron and a gamma quantum:

$$\delta^2_{fi} = (r_1\varepsilon_{iC} - 1) - \frac{r_1\varepsilon_{iC}E_i}{E_0 - E_{D(r_2)}}. \tag{87}$$

We emphasize that the ultrarelativistic parameter $\delta^2_{fi}$ must be positive $(\delta^2_{fi} \geq 0)$. Because of this, from expression (87) we obtain a more stringent condition than ratio (76) for the number of emitted photons of the wave at the first vertex:

$$r_1 \geq r_{1D}^{\min} = \lceil r_{D(r_2)} \rceil, \quad r_{D(r_2)} = \frac{1}{\varepsilon_{iC}} + \frac{2}{\varepsilon'_{iC}}\left[1 - \sqrt{1 - \frac{1}{r_2\varepsilon'_{iBW}}}\right]^{-1} \tag{88}$$

The ratio (88) is a condition for matching the number of emitted and absorbed photons of the wave at the first and second vertices of the channel D so that the general law of conservation of energy in the Breit-Wheeler process is fulfilled. At fixed initial energies of the electron and gamma quantum, as well as quantum parameters $\varepsilon_{iC}$ and $\varepsilon'_{iBW}$ ratios (87), (88) uniquely determine the dependence of the outgoing angles of the gamma quantum and the electron (see Fig. 7). At the same time, the values of the electron energy $E^{\mp}_{D(r_2)}$ (85) determine the interval of change in the outgoing angles for the final gamma quantum (compare with the ratio (75)):

$$\delta_-^2 \leq \delta_{fi}^2 \leq \delta_+^2, \qquad (89)$$

where

$$\delta_{\mp}^2 = (r_1 \varepsilon_{iC} - 1) - \frac{r_1 \varepsilon_{iC} E_i}{E_0 - E^{\pm}_{D(r_2)}}. \qquad (90)$$

Thus, the ratios (89), (90) determine the interval of change of the ultrarelativistic parameter $\delta_{fi}^2$ (the outgoing angle of the gamma quantum) in the ratio for the energy of the gamma quantum (73). Note that under conditions based on the energy of the initial particles (25), taking into account the ratios (78) and (86), we obtain the resonant energy of the final electron expressed in terms of the outgoing angle of the final gamma quantum and the number of emitted photons of the wave at the first vertex (88):

$$E_{D(r_1)} \approx \omega_i - \frac{1}{r_1}\left(1 + \delta_{fi}^2\right)\omega_C. \qquad (91)$$

It is important to note that the energies of the final particles for channels B and D lie in different energy regions. So, for channel B, the relations take place: $\omega_{B(r_1)} < E_i$, $E_{B(r_2)} > \omega_i$ (see expressions (38) and (40)). On the other hand, for channel D we have opposite inequalities: $\omega_{D(r_1)} > E_i$, $E_{D(r_2)} < \omega_i$ (see relations (73) and (79)). Thus, channels B and D are distinguishable and do not interfere.

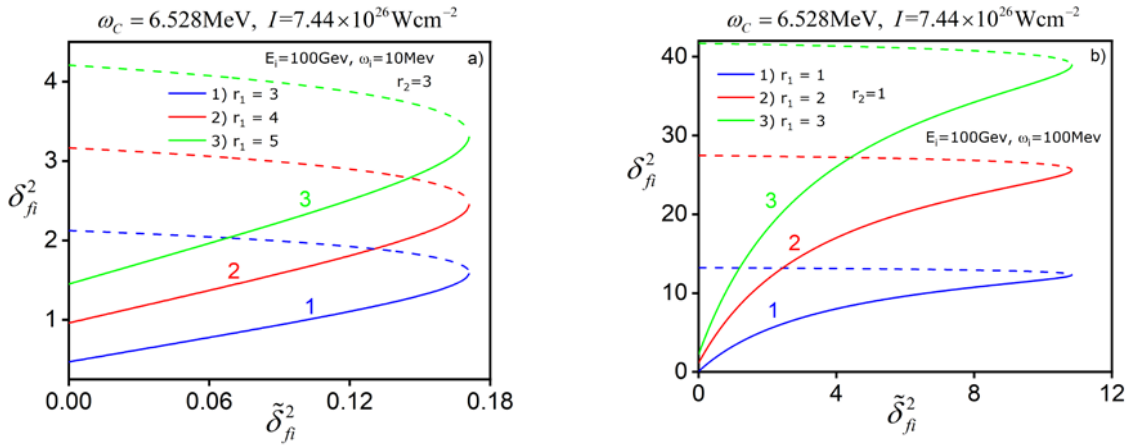

Fig.7. Dependence of ultrarelativistic parameters (87), (79), (81), (89) determining the outgoing angles of the gamma quantum and the electron, with a fixed characteristic energy of the Compton effect, the intensity of the X-ray wave and the energy of the initial electron, as well as for a different number of emitted and absorbed photons of the wave and different energies the initial gamma quantum. Solid (dashed) curves correspond to the "plus" ("minus") sign in front of the square root in the ratio (79). Case a) responds $r_{2D}^{min} = r_{1D}^{min} = 3$, case b) - $r_{2D}^{min} = r_{1D}^{min} = 1$.

The resonant differential cross-section for channel D can be obtained from expression (47) if in the latter we make reinterpretations (70) and take into account that at the first vertex we have an external field–stimulated inverse Breit-Wheeler process $\left(K_{r_1} \to -4P_{r_1}\right)$, and at the second vertex we have an external field-stimulated Breit-Wheeler process $\left(K_{r_2} \to -4P_{r_2}\right)$. After simple transformations, we get:

$$d\sigma_{D(r_1 r_2)} = r_e^2 \frac{2m^6 E_i}{\pi^2 \omega_i \omega_f E_f m_*^2 \left(1+\delta_i^2\right)} \frac{P_{r_1}\left(u_{1D}, v_{1D}\right) P_{r_2}\left(u_{2D}, v_{2D}\right)}{\left[\left(\tilde{q}_+^2 - m_*^2\right)^2 + 4m_*^2 \Gamma^2\left(\eta, \varepsilon'_{iBW}\right)\right]} \times \tag{92}$$

$$\times \delta^{(4)}\left[\tilde{p}_f + k_f - \tilde{p}_i - k_i - (r_2 - r_1)k\right] d^3\tilde{p}_f d^3 k_f.$$

Here, the functions $P_{r_1}(u_{1D}, v_{1D})$ and $P_{r_2}(u_{2D}, v_{2D})$ determining the probability of the external field-stimulated Breit-Wheeler process are determined by the expression [19]:

$$P_{r_j}\left(u_{jD}, v_{jD}\right) = J_{r_j}^2\left(\gamma_{jD}\right) + \eta^2 \left(2u_{jD} - 1\right)\left[\left(\frac{r_j^2}{\gamma_{jD}^2} - 1\right) J_{r_j}^2 + J_{r_j}'^2\right], \quad j = 1, 2 \tag{93}$$

At the same time, the arguments of the Bessel functions and the corresponding relativistically invariant parameters are defined by expressions:

$$\gamma_{jD} = 2r_j \frac{\eta}{\sqrt{1+\eta^2}} \sqrt{\frac{u_{jD}}{v_{jD}}\left(1 - \frac{u_{jD}}{v_{jD}}\right)}, \quad j = 1, 2, \tag{94}$$

$$u_{1D} = \frac{(kk_f)^2}{4(kq_+)(kp_i)} \approx \frac{\omega_f^2}{4E_i(\omega_f - E_i)}, \quad v_{1D} = r_1 \frac{(kk_f)}{2m_*^2} \approx r_1 \frac{\omega_f}{\omega_{BW}} \tag{95}$$

$$u_{2D} = \frac{(kk_i)^2}{4(kq_+)(kp_f)} \approx \frac{\omega_i^2}{4E_f(\omega_i - E_f)}, \quad v_{2D} = r_2 \frac{(kk_i)}{2m_*^2} \approx r_2 \varepsilon'_{BW} \tag{96}$$

The radiation width $\Gamma(\eta, \varepsilon'_{iBW})$ is determined by the expression (35), (32). By performing the corresponding integrations in expression (92), the resonant differential cross section for channel D under conditions (20)-(23) will take the following form:

$$R_{D(r_1 r_2)} = \frac{d\sigma_{D(r_1 r_2)}}{d\delta_{fi}^2} = \frac{2r_e^2}{\pi\left(1+\delta_i^2\right)} \left(\frac{m^4}{m_*^4}\right) \frac{m^2 E_i}{\omega_i \omega_f (E_0 - \omega_f)} \frac{P_{r_1}(u_{1D}, v_{1D}) P_{r_2}(u_{2D}, v_{2D})}{\left[\left(\delta_{fi(0)}^2 - \delta_{fi}^2\right)^2 + \Upsilon_{fi}'^2\right]} \tag{97}$$

Here, $\delta_{fi(0)}^2$ is the ultrarelativistic parameter, which determines the outgoing angle of the final gamma quantum relative to the momentum of the initial electron, changes independently of the resonant frequency of the gamma quantum, and the ultrarelativistic parameter $\delta_{fi}^2$ is related to the resonant frequency by the ratio (73). The value $\Upsilon_{fi}'$ is the angular resonant width of the channel D, equal to

$$\Upsilon_{fi}' = \frac{\alpha m^2}{4\pi m_*^2}\left(\frac{E_i}{\omega_f}\right) K\left(\eta, \varepsilon'_{iBW}\right) \tag{98}$$

When the condition is met

$$\left(\delta_{fi(0)}^2 - \delta_{fi}^2\right)^2 \ll \Upsilon_{fi}'^2, \tag{99}$$

after simple transformations, we obtain the maximum resonant cross section:

$$R_{D(r_1 r_2)}^{\max} = \frac{d\sigma_{D(r_1 r_2)}^{\max}}{d\delta_{fi}^2} = r_e^2 c_D \Psi_{D(r_1 r_2)}. \tag{100}$$

Here the function $c_D$ is determined by the initial installation parameters

$$c_D = \frac{8\pi}{\alpha^2 \left(1+\delta_i^2\right) \mathrm{K}^2\left(\eta, \varepsilon'_{iBW}\right)} \left(\frac{m}{\omega_C}\right)^2 \tag{101}$$

and the functions $\Psi_{D(rr')}$ determine the spectral-angular distribution of the resonant scattering cross section for channel D:

$$\Psi_{D(r_1 r_2)} = \frac{r_1 \mathrm{P}_{r_1}\left(u_{1D}, v_{1D}\right) \mathrm{P}_{r_2}\left(u_{2D}, v_{2D}\right)}{\varepsilon'_{iBW}\left[r_1 \varepsilon'_{iC} - \left(1+\delta_{fi}^2\right)\left(E_0/E_i\right)\right]}. \tag{102}$$

Here, the relativistically invariant parameters (95), as well as the argument of the Bessel functions $\gamma_{1D}$ (94) for $\omega_f = \omega_{D(r_1)}$ (73) take the following form:

$$u_{1D} \approx \frac{r_1 \varepsilon_{iC}}{4\left(1+\delta_{fi}^2\right)\xi_{fi(r_1)}}, \quad v_{1D} \approx \frac{r_1 \varepsilon_{iC}}{4\xi_{fi(r_1)}}. \tag{103}$$

$$\gamma_{1D} = 2r_1 \frac{\eta}{\sqrt{1+\eta^2}} \frac{\delta_{fi}}{\left(1+\delta_{fi}^2\right)}. \tag{104}$$

The function $\xi_{fi(r_1)}$ in the ratio (103) has the form (74). The relativistically invariant parameters (96), as well as the argument of the Bessel functions $\gamma_{2D}$ (94) for $E_f \approx E_{D(r_2)} \approx E_0 - \omega_{D(r_1)}$ take the following form:

$$u_{2D} \approx \frac{r_1 \varepsilon'_{iBW} \xi_{fi(r_1)}}{\left(1+\delta_{fi}^2\right)}\left[1 - \frac{\left(1+\delta_{fi}^2\right)}{r_1 \varepsilon'_{iC} \xi_{fi(r_1)}}\right]^{-1}, \quad v_{2D} \approx r_2 \varepsilon'_{iBW}. \tag{105}$$

Consider the case when the energy of the initial electron significantly exceeds the characteristic energy of the Compton effect (25). Then, the quantum parameter $\varepsilon_{iC} \gg 1$ and the resonant frequency of the final gamma quantum will be close to the energy of the initial electron (see expression (78)). At the same time, the parameters $u_{1D}$ and $v_{1D}$ (103) will be much larger than one:

$$u_{1D} \approx \frac{r_1 \varepsilon_{iC}}{4\left(1+\delta_{fi}^2\right)} \gg 1, \quad v_{1D} \approx \frac{1}{4} r_1 \varepsilon_{iC} \gg 1, \tag{106}$$

and the parameters $u_{2D}$ and $v_{2D}$ (105) take the form:

$$u_{2D} \approx \frac{r_1 \varepsilon'_{iBW}}{\alpha_{2D}}, \quad v_{2D} \approx r_2 \varepsilon'_{iBW}. \tag{107}$$

It is indicated here:

$$\alpha_{2D} = \left(1+\delta_{fi}^2\right)\left[1 - \frac{\left(1+\delta_{fi}^2\right)}{r_1 \varepsilon'_{iC}}\right]. \tag{108}$$

As a result, the expression for the function $\mathrm{P}_{r_1}\left(u_{1D}, v_{1D}\right)$ is significantly simplified and takes the form:

$$\mathrm{P}_{r_1}\left(u_{1D}, v_{1D}\right) \approx \frac{r_1 \eta^2 \varepsilon_{iC}}{4\left(1+\delta_{fi}^2\right)}\left[J_{r_1+1}^2 + J_{r_1-1}^2 - 2J_{r_1}^2\left(\gamma_{1D}\right)\right] \tag{109}$$

Here, the argument of the Bessel functions $\gamma_{1D}$ has the form (104). In this case, the argument of the Bessel function $\gamma_{2D}$ (94) in expression $\mathrm{P}_{r_2}\left(u_{2D}, v_{2D}\right)$ (93) takes the form:

$$\gamma_{2D} \approx 2r_1 \frac{\eta}{\alpha_{2D}\sqrt{1+\eta^2}} \sqrt{\frac{r_2}{r_1}\alpha_{2D} - 1}. \tag{110}$$

Considering this, after simple transformations, the expression for the maximum resonant section (100) will take the form:

$$R'^{max}_{D(r_1 r_2)} == r_e^2 c'_D \Psi'_{D(r_1 r_2)} \quad (E_i \gg \omega_C), \qquad (111)$$

where the functions $c'_D$ and $\Psi'_{D(r_1 r_2)}$ have the following form:

$$c'_D = \frac{2\pi \eta^2}{\alpha^2 (1+\delta_i^2)} \left(\frac{m}{\omega_C}\right)^2 \frac{\varepsilon_{iC}}{\varepsilon'_{iBW} K^2(\eta, \varepsilon'_{iBW})}, \qquad (112)$$

$$\Psi'_{D(r_1 r_2)} = \frac{r_1^2 P_{r_2}(u_{2D}, v_{2D})}{(1+\delta_{fi}^2)\left[r_1 \varepsilon'_{iC} - (1+\delta_{fi}^2)\right]} \left[J_{r_1+1}^2 + J_{r_1-1}^2 - 2J_{r_1}^2(\gamma_{1D})\right]. \qquad (113)$$

Here, the number of photons of the wave at the first vertex is determined by the expression (88), and at the second vertex by the expression (83). A comparison of the resonant sections (100)-(102) and (111)-(113) shows that the resonant differential cross section for the energies of the initial electron significantly exceeding the characteristic energy of the Compton effect (25) significantly (by a factor of $\varepsilon_{iC} \gg 1$) exceeds the corresponding resonant cross section when the energy of the initial electron is of the order of the characteristic energy of the Compton effect $(\varepsilon_{iC} \sim 1)$. At the same time, the parameter $\varepsilon'_{iBW}$ must be greater than one $(\omega_i > \omega_{BW})$. In the opposite case, when $\omega_i < \omega_{BW}$ the resonant cross section (111)-(113) is suppressed (see the conditions for the number of absorbed and emitted photons of the wave (83), (88)). Therefore, the estimate of the maximum resonant cross section $(r_1 = r_2 = 1)$ by function $c'_D$ (112) is valid only for $\varepsilon'_{iBW} \geq 1$ (for channel B, the corresponding estimate is valid for any parameter values $\varepsilon'_{iC}$). Note that at the first vertex a final gamma quantum is produced and the process is characterized by the characteristic energy of the Compton effect $\omega_C$, and at the second vertex a final electron is produced and the process is characterized by the characteristic Breit-Wheeler energy $\omega_{BW} = 4\omega_C$. So, for $\omega_C = 130.56$ MeV now we have $\omega_{BW} = 522,24$ MeV. Then, for the energies of the initial gamma quantum $\omega_i = 10$ MeV and $\omega_i = 100$ MeV, we obtain that the number of photons of the external field at the second vertex can take the following values: $r_2 \geq r_{2D}^{min} = 53$ and $r_2 \geq r_{2D}^{min} = 6$, respectively (see ratio (83)) At the same time, the number of photons in the first vertex is determined by the ratio (88). Therefore, for a given characteristic Breit-Wheeler energy, the resonant process will be suppressed $\left(R'^{max}_{D(r_1 r_2)} \ll r_e^2\right)$. For the same reason, the case should be excluded when $\omega_C = 26.112$ MeV, $\omega_{BW} = 104.448$ MeV, and $\omega_i = 10$ MeV. Figures 9, 10 show the dependences of the resonant cross section (111)-(113) on the square of the outgoing angle of the final gamma quantum for various energies of the initial particles, as well as various characteristic energies of the Compton effect and Breit-Wheeler, and the numbers of absorbed $(r_2)$ and emitted $(r_1)$ photons of the wave. Tables 4, 5 show the values of the maximum resonant cross section corresponding to the peaks in Figures 8, 9, as well as the corresponding values of the resonant energies of final gamma quanta and electrons. It can be seen from Figures 8 and 9 that under conditions where the parameter $\varepsilon_{iC} \gg 1$, the angular distribution of the resonant cross section significantly depends on the value of the parameter $\varepsilon'_{iBW}$. At the same time, the energy of the final gamma quantum always exceeds the energy of the initial electron $\left(\omega_{D(1,1)} > 100 \text{ GeV}\right)$. So, if $\varepsilon'_{iBW} \geq 1$, then $r_{2D}^{min} = r_{1D}^{min} = 1$ (see relations (83), (88)). In this case, the resonant cross section has a maximum value for the numbers of absorbed and emitted photons of the wave $r_1 = r_2 = 1$ (see Figure 8b) and is equal to $R_{D(1,1)}^{\max(*)} = 1.39 \times 10^5 r_e^2$. Moreover, there are two symmetric maxima for the outgoing angles of the final gamma quantum at the ends of the scattering angle interval (89): $\delta_{fi}^2 = \delta_-^2 = 0.07$ and $\delta_{fi}^2 = \delta_+^2 = 13.23$. With an increase in the number of emitted photons of the wave at the first vertex $(r_1 = 2,3)$, the resonant cross section decreases sharply, and the symmetry of the two maxima at the boundaries of the scattering angle interval disappears. At the same time, the value of the first resonant maximum $(\delta_{fi}^2 = \delta_-^2 = 1.15)$ 3.4 times exceeds the value of the second maximum $(\delta_{fi}^2 = \delta_+^2 = 27.46)$ (see Table 4). If it is a parameter $\varepsilon'_{iBW} < 1$, then the angular distribution of the resonant cross section changes significantly (see Fig.8a and Fig.9). So, if $r_{2D}^{\min} = r_{1D}^{\min} = 2$ (see Fig.9), then the maxima of the angular distribution shift from the edges into the interval of the outgoing angles of the final gamma quantum. At the same time, the value of the resonant cross section for the second maximum (for a larger outgoing angle) is always less than the value of the first maximum (for a smaller outgoing angle). The maximum resonant cross section takes place for $r_1 = r_2 = 2$ and is the value $R_{D(2,2)}^{\max(*)} = 1.62 \times 10^3 r_e^2$ for $\delta_{fi}^{2(*)} = 0.51$ and $R_{D(2,2)}^{\max(*)} = 1.36 \times 10^3 r_e^2$ for $\delta_{fi}^{2(*)} = 5.14$. With an increase in the number of emitted photons of the wave $r_1 = 3,4$, the resonant cross section decreases sharply, and the second maximum decreases most strongly (see Fig.9 and Table 5). If we reduce the characteristic energy of the Compton effect and the energy of the initial gamma quantum (increase the parameter $\varepsilon_{iC}$ and decrease the parameter $\varepsilon'_{iBW}$, see Fig.8a), then $r_{2D}^{\min} = r_{1D}^{\min} = 3$. In this case, the angular distribution curves of the resonant cross section have one maximum (the second maximum disappears), which decreases with an increase in the number of emitted photons of the wave $r_1 = 3,4,5$. So, for $r_1 = r_2 = 3$ the resonant cross section is the value $R_{D(3,3)}^{\max(*)} = 8.99 \times 10^4 r_e^2$ for $\delta_{fi}^{2(*)} = 1.27$. If $r_1 = 5, r_2 = 3$, then we get $R_{D(5,3)}^{\max(*)} = 5.93 \times 10^3 r_e^2$ for $\delta_{fi}^{2(*)} = 2.43$ (see Table 4).

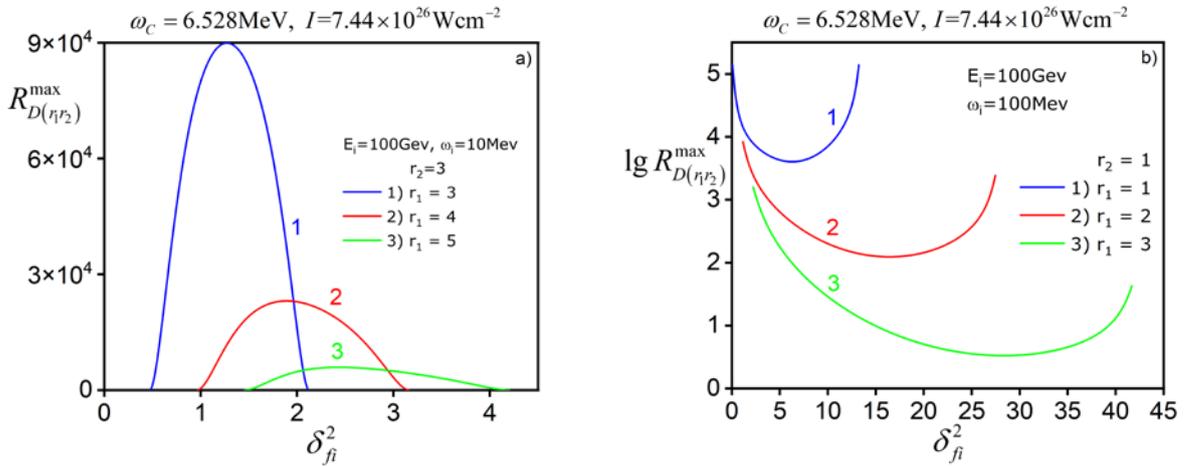

Figure 8. Dependence of the maximum resonant cross section (111) on the ultrarelativistic parameter $\delta_{fi}^2$ determining the outgoing angle of the gamma quantum for different energies of the initial gamma quantum and electron and the numbers of emitted and absorbed photons of the wave at fixed parameters of a strong X-ray wave. Case a) responds $r_{2D}^{\min} = r_{1D}^{\min} = 3$, case b) - $r_{2D}^{\min} = r_{1D}^{\min} = 1$.

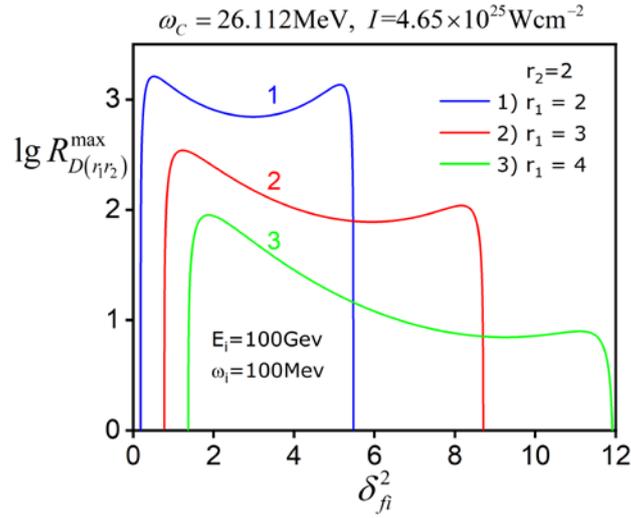

Fig. 9. Dependence of the maximum resonant cross section (111) on the ultrarelativistic parameter $\delta_{fi}^2$ determining the outgoing angle of the gamma quantum for various numbers of emitted and absorbed photons of the wave and at fixed values of the initial energies of the gamma quantum and electron, as well as the parameters of a strong X-ray wave. Here $r_{2D}^{min} = r_{1D}^{min} = 2$.

Table 4.

$\omega_C = 6.528$ MeV, $\omega = 20$ keV, $I = 7.44 \times 10^{26}$ Wcm$^{-2}$

| (GeV) | $r_1, r_2$ | $\delta_{fi}^{2(*)}$ | $\omega_D$ (GeV) | $\tilde{\delta}_{fi}^{2(*)}$ | $E_D$ (MeV) | $R_{D(r_1 r_2)}^{max(*)}$ $(r_e^2)$ |
|---|---|---|---|---|---|---|
| | 1, 1 | 0.07 | 100.007021 | 0 | 92.979 | $1.39 \times 10^5$ |
| | 1, 1 | 13.23 | 100.092979 | 0 | 7.021 | $1.39 \times 10^5$ |
| $E_i = 100$ | 2, 1 | 1.15 | 100.007021 | 0 | 92.979 | $8.40 \times 10^3$ |
| $\omega_i = 0.1$ | 2, 1 | 27.46 | 100.092979 | 0 | 7.021 | $2.46 \times 10^3$ |
| | 3, 1 | 2.22 | 100.007021 | 0 | 92.979 | $1.58 \times 10^3$ |
| | 3, 1 | 41.69 | 100.092980 | 0 | 7.020 | 42.93 |
| $E_i = 100$ | 3, 3 | 1.27 | 100.004943 | 0.15 | 5.057 | $8.99 \times 10^4$ |
| $\omega_i = 0.01$ | 4, 3 | 1.89 | 100.004720 | 0.13 | 5.280 | $2.31 \times 10^4$ |
| | 5, 3 | 2.43 | 100.004481 | 0.11 | 5.519 | $5.93 \times 10^3$ |

Table 5.

$\omega_C = 26.112$ MeV, $\omega = 5$ keV, $I = 4.65 \times 10^{25}$ Wcm$^{-2}$

| (GeV) | $r_1, r_2$ | $\delta_{fi}^{2(*)}$ | $\omega_D$ (GeV) | $\tilde{\delta}_{fi}^{2(*)}$ | $E_D$ (MeV) | $R_{D(r_1 r_2)}^{max(*)}$ $(r_e^2)$ |
|---|---|---|---|---|---|---|
| | 2, 2 | 0.51 | 100.01975 | 0.08 | 80.250 | $1.62 \times 10^3$ |
| | 2, 2 | 5.14 | 100.080273 | 1.37 | 19.727 | $1.36 \times 10^3$ |
| $E_i = 100$ | 3, 2 | 1.23 | 100.019393 | 0.08 | 80.607 | $3.46 \times 10^2$ |
| $\omega_i = 0.1$ | 3, 2 | 8.16 | 100.079815 | 1.44 | 20.185 | $1.09 \times 10^2$ |
| | 4, 2 | 1.87 | 100.018752 | 0.06 | 81.248 | $8.97 \times 10^1$ |
| | 4, 2 | 11.11 | 100.079117 | 1.52 | 20.883 | 7.93 |

## 5. Suppression of channel A for high energies of initial electrons

For channel A, the laws of conservation of the 4-momentum at the first and second vertices can be represented as follows:

$$\tilde{p}_i + k_i = \tilde{q}_- + r_1 k, \tag{114}$$

$$\tilde{q}_- + r_2 k = \tilde{p}_f + k_f. \tag{115}$$

Since $\tilde{p}_{i,f}^2 = \tilde{q}_-^2 = m_*^2$ than equations (114), (115) are valid only for $r_1 \geq 1$ and $r_2 \geq 1$. Hence, it follows from the type of amplitude (6)-(8) (see also Fig.2) that under resonant conditions, the second-order process by the fine structure constant is effectively reduced to two first-order processes of the external field-stimulated Compton effect. At the first vertex, we have the scattering of the initial gamma quantum on the initial electron with the emission of an intermediate electron and $r_1$-photons of the wave. At the second vertex, we have the absorption by an intermediate electron $r_2$-photons of a wave with the emission of a final gamma quantum and an electron.

From the ratios (27), (114), taking into account expressions (20)-(23), after simple transformations, we obtain a condition for the angle of solution between the momenta of the initial particles:

$$\delta_i^2 = \delta_{res(r_1)}^2, \quad \delta_{res(r_1)}^2 = (r_1 \varepsilon_{*C} - 1) \geq 0 \tag{116}$$

where is indicated

$$\varepsilon_{*C} = \frac{E_*}{\omega_C}, \quad \delta_i^2 = \frac{E_i^2 \theta_i^2}{m_*^2}, \tag{117}$$

$$E_* = \frac{E_0 E_i}{\omega_i}, \quad E_0 = E_i + \omega_i. \tag{118}$$

Here $E_*$ and $E_0$ are the combination energy and the total energy of the initial particles. The inequality follows from the ratio (116):

$$r_1 \varepsilon_{*C} \geq 1 \tag{119}$$

Because of this, for the resonant process of channel A, the number of emitted photons of the wave $r_1$ significantly depends on the magnitude of the quantum parameter $\varepsilon_{*C}$. So, if the combinational energy of the initial particles is less than the characteristic energy $(E_* < \omega_C)$, then the quantum parameter $\varepsilon_{*C} < 1$. In this case, there is a minimum number of emitted photons of the wave $r_{1 \min}$, starting from which the resonant process takes place:

$$r_1 \geq r_{1A}^{\min} = \lceil \varepsilon_{*C}^{-1} \rceil \quad (E_* < \omega_C) \tag{120}$$

If the combinational energy of the initial particles is greater than or equal to the characteristic energy $(E_* \geq \omega_C)$, then the quantum parameter $\varepsilon_{*C} \geq 1$. Therefore, in this case, the resonance of channel A takes place when studying one or more photons of the wave

$$r_1 \geq 1 \quad (E_* \geq \omega_C) \tag{121}$$

Thus, for channel A, unlike channels B and D, the resonant condition at the first vertex imposes a strict condition on the ultrarelativistic parameter $\delta_i^2$ that determines the initial angle of the solution between the momenta of the initial particles.

Note that the resonant condition for the second vertex of channel A (27), (115) allows us to obtain a cubic equation for the resonant frequency of a final gamma quantum [56]. The resonant differential cross section for channel A is easy to obtain using the corresponding results for channels B and D. After simple calculations, we obtain an estimate of the order of magnitude of the maximum resonant differential cross section for channel A:

$$R_{A(r_1 r_2)}^{\max} = \frac{d\sigma_{A(r_1 r_2)}^{\max}}{d\delta_f^2} \lesssim c_A r_e^2, \qquad (122)$$

Here the function $c_A$ is determined by the initial installation parameters

$$c_A = \frac{2\pi}{\alpha^2 \mathrm{K}^2(\eta, \varepsilon_{0C})} \left(\frac{m}{E_0}\right)^2 \lesssim 10^5 \left(\frac{m}{E_0}\right)^2. \qquad (123)$$

Expressions (122) and (123) denote:

$$\delta_f^2 = \frac{E_0^2 \theta_f^2}{m_*^2}, \quad \varepsilon_{0C} = \frac{E_0}{\omega_C}. \qquad (124)$$

In this case, the resonant width $\mathrm{K}(\eta, \varepsilon_{0C})$ is determined by the expression (32). Let's estimate the order of magnitude of the resonant cross section (122), (123). For the energies of the initial particles (24), we obtain

$$R_{A(r_1 r_2)}^{\max} \lesssim 10^{-2} r_e^2. \qquad (125)$$

Thus, for the energies of the initial particles studied in this article, channel A is suppressed relative to channels B and D. In addition, in this article we studied the resonance of channels B and D under conditions when the resonant condition at the first vertex for the initial particles (116) was not fulfilled. Thus, the resonance of channel A did not take place.

In conclusion of this section, we note that the corresponding nonresonant differential cross section of the Compton effect in this geometry (20)-(22) has the following order of magnitude [29]

$$R_{nonres} = \frac{d\sigma_{nonres}}{d\delta_f^2} \sim r_e^2. \qquad (126)$$

## 6. Results and discussions

The study of the resonant inverse Compton effect in the field of a strong electromagnetic wave in the region of high energies of the initial electrons (24), (25) showed that in this energy region the main reaction channels are the scattering channel B and the annihilation channel D. In this case, the resonant channel A is suppressed.

The resonant process is characterized by two quantum parameters: the characteristic energy of the Compton effect $\omega_C$ and the characteristic Breit-Wheeler energy $\omega_{BW}$ (3), which is four times higher than the characteristic energy of the Compton effect. These quantum energies characterize the corresponding first-order QED process stimulated by an external field [19].

The resonant process is characterized by an unambiguous dependence of the outgoing angles of the final particles (see ratios (43) for channel B and (87) for channel D, as well as Fig.3 and Fig.7). This dependence is a distinctive feature of the resonant process, in contrast to the non-resonant process, for which the outgoing angles of the final particles are independent of each other.

The energies of the final particles and the magnitude of the resonant differential cross section are determined by the outgoing angle of the final gamma quantum and quantum parameters equal to the ratio of the initial energy of the particle to the corresponding characteristic energy (2). Thus, the scattering channel B is characterized by quantum parameters $\varepsilon_{iC}$ (at the first vertex) and $\varepsilon'_{iC}$ (in the second vertex). The annihilation channel D is characterized by quantum parameters $\varepsilon_{iC}$ (at the first vertex) and $\varepsilon'_{iBW}$ (in the second vertex). It is the magnitude of these quantum parameters that determines which resonant channel will prevail.

A final gamma quantum is generated at the first vertex of channels B and D, and a final electron is generated at the second vertex. At the same time, for the channel B the energy of final particles, the following relations are satisfied: $\omega_{B(r_1)} < E_i$, $E_{B(r_2)} > \omega_i$; and for the channel D we have the opposite inequalities: $\omega_{D(r_1)} > E_i$, $E_{D(r_2)} < \omega_i$. Thus, the B and D channels are energetically distinguishable and do not interfere.

The resonant energy of a final gamma quantum depends significantly on the magnitude of the quantum parameter $\varepsilon_{iC}$. If the parameter $\varepsilon_{iC} \gg 1$ (the energy of the initial electron significantly exceeds the characteristic energy of the Compton effect), then the energy of the final gamma quantum tends to the energy of the initial electron, and for channel B from below $\left(E_i - \omega_{B(r_1)}\right) \ll 1$, and for channel D from above $\left(\omega_{D(r_1)} - E_i\right) \ll 1$. Thus, in this case we have the inverse Compton effect, i.e., the energy of the initial high-energy electrons is converted into the energy of the final gamma quanta (see Tables 1-5).

The magnitude of the resonant cross section is significantly affected by the number of photons absorbed and emitted at the first and second vertices of the wave. At the same time, for the scattering channel, the number of absorbed photons of the wave at the first vertex can be arbitrary $(r_1 \geq 1)$, and the number of emitted photons at the second vertex is limited from below by the condition (63). This limitation on the number of emitted photons is due to the coordination of resonant processes in the first and second vertices of channel B by virtue of the general law of conservation of energy. For the annihilation channel D, the number of absorbed photons of the wave at the second vertex is limited by the ratios (83), (84). This is due to the fact that the external field-stimulated Breit-Wheeler process has a reaction threshold. The number of emitted photons of the wave at the first vertex is limited by the ratio (88). This limitation on the number of emitted photons is due to the coordination of resonant processes in the first and second vertices of the channel D by virtue of the general law of conservation of energy.

For an X-ray wave $(\omega = 1\,\text{keV})$ with intensity $I = 1.86 \cdot 10^{24}\,\text{Wcm}^{-2}$ and characteristic energies $\omega_C \approx 130.56\,\text{MeV}$ and $\omega_{WC} \approx 522.24\,\text{MeV}$ the annihilation channel D is suppressed ( $R_{D(r_1 r_2)}^{\max} \ll r_e^2$ ) for the initial energies of the gamma quantum $\omega_i = 10\,\text{MeV}$ and $\omega_i = 100\,\text{MeV}$. In this case, high-energy gamma quanta $\left(\omega_{B(r_1)} \approx E_i \gg \omega_C\right)$ are most likely generated through channel B, and most likely when the energy of the initial gamma quantum is much less than the characteristic energy of the Compton effect. At the same time, the resonant cross section takes the maximum value for the number of absorbed and emitted photons of the wave $r_1 = r_2 = 1$. So, for the initial energies of the electron $E_i = 100\,\text{GeV}$ and gamma quanta $\omega_i = 10\,\text{MeV}$, and $\omega_i = 100\,\text{MeV}$ the maximum resonant cross section takes the following values $R_{B(1,1)}^{\max(*)} \approx 2.68 \times 10^2\, r_e^2$ and $R_{B(1,1)}^{\max(*)} \approx 8.86\, r_e^2$, respectively. In this case, the final gamma quantum is emitted along the momentum of the initial electron $\left(\delta_{fi}^{2(*)} = 0\right)$ with energy $\omega_{B(1)} \approx 99.87\,\text{GeV}$ (see Fig.4 and Table 1). If the initial energies of the particles are equal to $E_i = 1\,\text{GeV}$ and $\omega_i = 10\,\text{MeV}$, then the resonant cross section takes on a value $R_{B(1,1)}^{\max(*)} \approx 22.91\, r_e^2$ and the gamma quantum is emitted with energy $\omega_{B(1)} \approx 0.88\,\text{GeV}$.

If we increase the intensity of the X-ray wave $(\omega = 5\,\text{keV})$ to a value $I = 4.65 \cdot 10^{25}\,\text{Wcm}^{-2}$ (reduce the characteristic energies to values $\omega_C \approx 26.112\,\text{MeV}$, $\omega_{BW} \approx 104.448\,\text{MeV}$ ), then along with channel B, at energies of the initial gamma quantum comparable to the characteristic Breit-Wheeler energy, channel D begins to appear. However, for the initial energies of the particles $E_i = 100\,\text{GeV}$, $\omega_i = 10\,\text{MeV}$, the D channel will be suppressed ( $R_{D(r_1 r_2)}^{\max} \ll r_e^2$ ), since in this case the quantum parameter $\varepsilon'_{iBW} \approx 0.096 \ll 1$. In this case, the main resonant channel is channel B. Through this channel, high-energy gamma quanta $\left(\omega_{B(1)} \approx 99.974 \text{ GeV}\right)$ with a resonant cross section $R_{B(1,1)}^{\max(*)} \approx 2.09 \times 10^3 \, r_e^2$ will be generated along the momentum of the initial electron $\left(\delta_{fi}^{2(*)} = 0\right)$ (see Fig.5 and Table 2). However, if the energy of the initial gamma quantum is increased tenfold $\left(\omega_i = 100 \text{ MeV}\right)$, the annihilation channel D becomes the main one, the resonant cross section of which exceeds the corresponding cross section of channel B by an order of magnitude. Indeed, if the energies of the initial particles are equal to $E_i = 100 \text{ GeV}$ and $\omega_i = 100 \text{ MeV}$, then the channel D for a fixed number of absorbed and emitted photons of the wave has two maxima (see Fig. 9). At the same time, the maximum resonant sections $(r_1 = r_2 = 2)$, their corresponding outgoing angles and the energy of the gamma quantum take the following values: $R_{D(2,2)}^{\max(*)} \approx 1.62 \times 10^3 \, r_e^2$, $\delta_{fi}^{2(*)} = 0.51$, $\omega_{D(2,2)} \approx 100.02 \text{ GeV}$ and $R_{D(2,2)}^{\max(*)} \approx 1.36 \times 10^3 \, r_e^2$, $\delta_{fi}^{2(*)} = 5.14$, $\omega_{D(2,2)} \approx 100.08 \text{ GeV}$. At the same time, through the channel B, the maximum resonant cross section takes place at $r_1 = r_2 = 1$ and is the value of $R_{B(1,1)}^{\max(*)} \approx 1.97 \times 10^2 \, r_e^2$. In this case, the gamma quantum flies out along the momentum of the initial electron $\left(\delta_{fi}^{2(*)} = 0\right)$ with energy $\omega_{B(1,1)} \approx 99.97 \text{ GeV}$.

When the intensity of the X-ray wave $\left(\omega = 20 \text{ keV}\right)$ increases to a value $I = 7.44 \cdot 10^{26} \text{ Wcm}^{-2}$ (a corresponding decrease in the characteristic energies to values $\omega_C \approx 6.528 \text{ MeV}$, $\omega_{BW} \approx 26.112 \text{ MeV}$), the annihilation channel D becomes predominant. The resonant cross-section for this channel significantly exceeds the corresponding resonant cross-section for channel B (see Fig.6, Fig.8 and Tab.3, Tab.4). Thus, for the initial energies of particles $E_i = 100 \text{ GeV}$, $\omega_i = 10 \text{ MeV}$, for channels D and B, the maximum resonant cross-sections, their corresponding outgoing angles and gamma-ray energies take the following values are: $R_{D(3,3)}^{\max(*)} \approx 8.99 \times 10^4 \, r_e^2$, $\delta_{fi}^{2(*)} = 1.27$, $\omega_{D(3,3)} \approx 100.005 \text{ GeV}$ and $R_{B(1,1)}^{\max(*)} \approx 1.76 \times 10^4 \, r_e^2$, $\delta_{fi}^{2(*)} = 0$, $\omega_{B(1,1)} \approx 99.993 \text{ GeV}$. At the same time, for the initial energies of the particles $E_i = 100 \text{ GeV}$, $\omega_i = 100 \text{ MeV}$, the resonant cross section for channel D is two orders of magnitude higher than the corresponding cross section of channel B. So, for channels D (here there are two maxima for different outgoing angles) and B, the maximum resonant cross sections, their corresponding outgoing angles and gamma quantum energy take the following values: $R_{D(1,1)}^{\max(*)} \approx 1.39 \times 10^5 \, r_e^2$, $\delta_{fi}^{2(*)} = 0.07 (13.23)$, $\omega_{D(1,1)} \approx 100.007 (100.09) \text{ GeV}$ and $R_{B(1,1)}^{\max(*)} \approx 3.13 \times 10^3 \, r_e^2$, $\delta_{fi}^{2(*)} = 0$, $\omega_{B(1,1)} \approx 99.993 \text{ GeV}$.

It is important to note that Oleinik resonances occur not only in the field of a plane monochromatic wave, but also in the field of a plane pulse wave, provided that the pulse time $\tau$ significantly exceeds the period of wave oscillations $\tau \gg \omega^{-1}$ [24, 30, 31]. However, for very short pulses, when $\tau \sim \omega^{-1}$ Oleinik resonances may not manifest. In this article, an idealized case of a plane monochromatic electromagnetic wave is considered. In a real experiment, as well as near pulsars and magnetars, the electromagnetic wave is inhomogeneous in space and time. The study of Oleinik resonances in such fields is a rather complex independent task that can be performed only by numerical solution of the corresponding mathematical problem. The solution of the resonant problem in the field of a plane monochromatic wave nevertheless allows solving a number of important problems. First, to identify the main physical parameters of the problem (the characteristic energy of the process (3), the quantum parameters (2)), which determine the resonant energy of the final particles, as well as the magnitude of the resonant differential cross section.

Secondly, it allowed us to obtain analytical expressions for the resonant differential scattering cross section. Note that all this is very important for the subsequent numerical analysis of the corresponding process in an inhomogeneous electromagnetic field.

Note also that we are considering sufficiently large energies of the initial electrons. Currently, obtaining narrow beams of ultrarelativistic electrons of such high energies at modern experimental facilities is problematic. However, in the Universe, in particular, near pulsars and magnetars, such energies of high-energy electrons and gamma quanta are possible. At the same time, near such objects in strong X-ray fields, cascades of resonant QED processes are possible, such as: resonant spontaneous bremsstrahlung during scattering of ultrarelativistic electrons on nuclei [56, 59], the Bethe-Heitler resonant process [54, 55, 58], the Breit-Wheeler resonant process [59], the resonant Compton-effect, etc. These processes are interconnected and can generate streams of high-energy gamma quanta and ultrarelativistic electrons and positrons. Thus, the results obtained can be used to explain narrow fluxes of high-energy gamma quanta near neutron stars, such as double X-ray systems operating on accretion [88, 89], X-ray/gamma pulsars operating on rotation [90, 91] and magnetars operating on a magnetic field [92, 93].

**Acknowledgments:**
The research is partially funded by the Ministry of Science and Higher Education of the Russian Federation as part of the World-class Research Center program: Advanced Digital Technologies (contract No. 075-15-2022-311, 20.04.2022)

**REFERENCES**

[1] Bula, C.; McDonald, K.T.; Prebys, E.J.; Bamber, C.; Boege, S.; Kotseroglou, T.; Melissinos, A.C.; Meyerhofer, D.D.; Ragg, W.; Burke, D.L.; et al., "Observation of Nonlinear Effects in Compton Scattering", Phys. Rev. Lett., 76, 3116 (1996).

[2] C. N. Danson, C. Haefner, J. Bromage, T. Butcher, J.-C. F. Chanteloup, E. A. Chowdhury, A. Galvanauskas, L. A. Gizzi, J. Hein, D. I. Hillier, and e. al, "Petawatt and exawatt class lasers worldwide," High Power Laser Science and Engineering 7, e54 (2019).

[3] J. W. Yoon, Y. G. Kim, I. W. Choi, J. H. Sung, H. W. Lee, S. K. Lee, and C. H. Nam, "Realization of laser intensity over 1023 w/cm2," Optica 8, 630–635 (2021).

[4] I. C. E. Turcu, B. Shen, D. Neely, G. Sarri, K. A. Tanaka, P. McKenna, S. P. D. Mangles, T.-P. Yu, W. Luo, X.-L. Zhu, and e. al, "Quantum electrodynamics experiments with colliding petawatt laser pulses," High Power Laser Science and Engineering 7, e10 (2019).

[5] K. A. Tanaka, K. M. Spohr, D. L. Balabanski, S. Balascuta, L. Capponi, M. O. Cernaianu, M. Cuciuc, A. Cucoanes, I. Dancus, A. Dhal, B. Diaconescu, D. Doria, P. Ghenuche, D. G. Ghita, S. Kisyov, V. Nastasa, J. F. Ong, F. Rotaru, D. Sangwan, P.-A. S¨oderstr¨om, D. Stutman, G. Suliman, O. Tesileanu, L. Tudor, N. Tsoneva, C. A. Ur, D. Ursescu, and N. V. Zamfir, "Current status and highlights of the ELI-NP research program," Matter and Radiation at Extremes 5, 024402 (2020).

[6] S. Weber, S. Bechet, S. Borneis, L. Brabec, M. Buˇcka, E. Chacon-Golcher, M. Ciappina, M. DeMarco, A. Fajstavr, K. Falk, , et al., "P3: An installation for high-energy density plasma physics and ultra-high intensity laser–matter interaction at eli-beamlines," Matter and Radiation at Extremes 2, 149–176 (2017).

[7] D. N. Papadopoulos, J. P. Zou, C. L. Blanc, G. Ch´eriaux, P. Georges, F. Druon, G. Mennerat, P. Ramirez, L. Martin, A. Fr´eneaux, and e. al, "The apollon 10 PW laser: experimental and theoretical investigation of the temporal characteristics," High Power Laser Science and Engineering 4, e34 (2016).

[8] J. Bromage, S.-W. Bahk, I. A. Begishev, C. Dorrer, M. J. Guardalben, B. N. Hoffman, J. B. Oliver, R. G. Roides, E. M. Schiesser, M. J. S. III, and e. al, "Technology development for ultraintense all-opcpa systems," High Power Laser Science and Engineering 7, e4 (2019).

[9] J. Rossbach, J. R. Schneider, and W. Wurth, "10 years of pioneering x-ray science at the free-electron laser FLASH at DESY," Physics Reports 808, 1–74 (2019).

[10] A. Gonoskov, A. Bashinov, S. Bastrakov, E. Efimenko, A. Ilderton, A. Kim, M. Marklund, I. Meyerov, A. Muraviev, and A. Sergeev, "Ultrabright GeV photon source via controlled electromagnetic cascades in laser-dipole waves," Phys. Rev. X 7, 041003 (2017).

[11] J. Magnusson, A. Gonoskov, M. Marklund, T. Z. Esirkepov, J. K. Koga, K. Kondo, M. Kando, S. V. Bulanov, G. Korn, and S. S. Bulanov, "Laser particle collider for multi-GeV photon production," Phys. Rev. Lett. 122, 254801 (2019).

[12] X.-L. Zhu, T.-P. Yu, M. Chen, S.-M. Weng, and Z.-M. Sheng, "Generation of GeV positron and -photon beams with controllable angular momentum by intense lasers," New Journal of Physics 20, 83013 (2018).

[13] P. Musumeci, C. Boffo, S. S. Bulanov, I. Chaikovska, A. F. Golfe, S. Gessner, J. Grames, R. Hessami, Y. Ivanyushenkov, A. Lankford, G. Loisch, G. Moortgat-Pick, S. Nagaitsev, S. Riemann, P. Sievers, C. Tenholt, and K. Yokoya, "Positron sources for future high energy physics colliders," arXiv:2204.13245 (2022).

[14] A. Alejo, G.M. Samarin, J.R. Warwick and G. Sarri, "Laser-Wakefield Electron Beams as Drivers of High-Quality Positron Beams and Inverse-Compton-Scattered Photon Beams", Front. Phys. 7:49 (2019).

[15] T. Y. Long et. al., "All-optical generation of petawatt gamma radiation via inverse Compton scattering from laser interaction with tube target", Plasma Phys. Control. Fusion 61, 085002 (2019).

[16] B. Terzić, J. McKaig, E. Johnson, T. Dharanikota, and G.A. Krafft, "Laser chirping in inverse Compton sources at high electron beam energies and high laser intensities", Phys. Rev. Accelerators and Beams 24, 094401 (2021).

[17] B.S. Günther. "Overview on Inverse Compton X-ray Sources. In: Storage Ring-Based Inverse Compton X-ray Sources", Springer Theses. Springer, Cham. (2023).

[18] F. V. Bunkin and M. V. Fedorov, "Bremsstrahlung in a strong radiation field," Sov. Phys. JETP 22, 844 (1966).

[19] V. Ritus and A. I. Nikishov, "Quantum Electrodynamics Phenomena in the Intense Field", Vol. 111 (Nauka, 1979).

[20] S.P. Roshchupkin, "Resonant effects in collisions of relativistic electrons in the field of a light wave," Laser Physics 6, 837–858 (1996).

[21] S.P. Roshchupkin, V.A. Tsybul'nik, A.N. Chmirev, "Probability of multiphoton processes in phenomena of a quantum electrodynamics in a strong light field Laser Phys.", 10, 1256 (2000).

[22] F. Ehlotzky, K. Krajewska, and J. Z. Kamínski, "Fundamental processes of quantum electrodynamics in laser fields of relativistic power," Reports on Progress in Physics 72, 46401 (2009).

[23] A. D. Piazza, C. Müller, K. Z. Hatsagortsyan, and C. H. Keitel, "Extremely high-intensity laser interactions with fundamental quantum systems," Rev. Mod. Phys. 84, 1177–1228 (2012).

[24] S. P. Roshchupkin, A. A. Lebed', E. A. Padusenko, and A. I. Voroshilo, "Quantum electrodynamics resonances in a pulsed laser field," Laser Physics 22, 1113 (2012).

[25] A. A. Mironov, S. Meuren, A. M. Fedotov, "Resummation of QED radiative corrections in a strong constant crossed field", Phys. Rev. D, 102(5), 053005, 18 (2020).

[26] A. Gonoskov, T. G. Blackburn, M. Marklund, S. S. Bulanov, "Charged particle motion and radiation in strong electromagnetic fields", Rev. Mod. Phys., 94(4), 045001, 63 (2022).

[27] A. Fedotov et al., "Advances in QED with intense background fields", Physics Reports, Volume 1010, 13 (2023).

[28] M. V. Fedorov, "An Electron in a Strong Light Field" (Nauka, Moscow) (1991).

[29]   S.P. Roshchupkin, A.I. Voroshilo, "Resonant and Coherent Effects of Quantum Electrodynamics in the Light Field" (Naukova Dumka: Kiev, Ukraine) (2008).

[30]   S. P. Roshchupkin, A. A. Lebed', E. A. Padusenko, and A. I. Voroshilo, "Resonant effects of quantum electrodynamics in the pulsed light field", in Quantum Optics and Laser Experiments, edited by S. Lyagushyn (Intech, Rijeka, Croatia) Chap. 6, (2012).

[31]   S. P. Roshchupkin, A. A. Lebed', "Effects of Quantum Electrodynamics in the Strong Pulsed Laser Fields" (Naukova Dumka: Kiev, Ukraine) (2013).

[32]   V. P. Oleinik, "Resonance effects in the field of an intense laser beam," Journal of Experimental and Theoretical Physics 25, 697 (1967).

[33]   V. P. Oleinik, "Resonance effects in the field of an intense laser ray," Journal of Experimental and Theoretical Physics 26, 1132 (1968).

[34]   N. B. Narozhny and M. S. Fofanov, "Photon emission by an electron in a collision with a short focused laser pulse", Zh. Eksp. Teor. Fiz. 110, 26 (1996).

[35]   N. B. Narozhny and M. S. Fofanov, "Scattering of relativistic electrons by a focused laser pulse", Zh. Eksp. Teor. Fiz. 90, 753 (2000).

[36]   M. Boca and V. Florescu, "Nonlinear Compton scattering with a laser pulse", Phys. Rev. A 80, 053403 (2009).

[37]   T. Heinzl, C. Harvey and A.Ilderton, "Signatures of high-intensity Compton scattering", Phys. Rev. A 79, 063407 (2009).

[38]   F. Mackenroth and A. Di Piazza, "Nonlinear Compton scattering in ultrashort laser pulses", Phys. Rev. A 83, 032106 (2011).

[39]   F. Mackenroth and A. Di Piazza, "Nonlinear Double Compton Scattering in the Ultrarelativistic Quantum Regime", Phys. Rev. Lett. 110, 070402 (2013).

[40]   D. Seipt and B. Kämpfer, "Two-photon Compton process in pulsed intense laser fields", Phys. Rev. D 85, 101701 (2012).

[41]   D. Seipt and B. Kampfer, "Nonlinear Compton scattering of ultrashort intense laser pulses", Phys. Rev. A 83, 022101 (2011).

[42]   V. Dinu, M. Boca and V. Florescu, "Electron distributions in nonlinear Compton scattering", Phys. Rev. A 86, 013414 (2012).

[43]   Ya-Nan Dai, Jing-Jing Jiang, Yu-Hang Jiang, Rashid Shaisultanov, and Yue-Yue Chen, "Effects of angular spread in nonlinear Compton scattering", Phys. Rev. D 108, 056025 (2023)

[44]   Khalaf and Kaminer, "Compton scattering driven by intense quantum light", Science Advances 9, eade0932 (2023).

[45]   A.I. Voroshilo and S.P. Roshchupkin, "Resonant scattering of a photon by an electron in the field of a circularly polarized electromagnetic wave", Laser Phys. Lett. 2, 184 (2005).

[46]   A. Voroshilo, S. Roshchupkin & O. Denisenko, "Resonance of exchange amplitude of Compton effect in the circularly polarized laser field", Eur. Phys. J. D 41, 433 (2007).

[47]   A.I. Voroshilo, S.P. Roshchupkin & V.N. Nedoreshta, "Resonant scattering of photon by electron in the presence of the pulsed laser field", Laser Phys. 21, 1675 (2011).

[48]   V. N. Nedoreshta, A. I. Voroshilo, and S. P. Roshchupkin, "Resonant scattering of a photon by an electron in the moderately-strong-pulsed laser field", Phys. Rev. A 88, 052109 (2013).

[49]   V. N. Nedoreshta, S. P. Roshchupkin, and A. I. Voroshilo, "Resonance of the exchange amplitude of a photon by an electron scattering in a pulsed laser field", Phys. Rev. A 91, 062110 (2015).

[50]   J. Bos, W. Brock, H. Mitter, and T. Schott, "Resonances and intensity dependent shifts of the moller cross section in a strong laser field," Journal of Physics A: Mathematical and General 12, 715–731 (1979).

[51]   A.V. Borisov, V. C. Zhukovskii, and P. A. ´Eminov, "Resonance bremsstrahlung of an electron by a nucleus in the field of a plane electromagnetic wave," Soviet Physics Journal 23, 184–188 (1980).

[52] A. V. Borisov, V. C. Zhukovskii, A. K. Nasirov, and P. A. ´Eminov, "Resonance two-photon pair production on nuclei and on electrons," Soviet Physics Journal 24, 107–110 (1981).

[53] K. Krajewska, "Electron-positron pair creation and Oleinik resonances," Laser Physics 21, 1275–1287 (2011).

[54] N. R. Larin, V. V. Dubov, and S. P. Roshchupkin, "Resonant photoproduction of high-energy electron-positron pairs in the field of a nucleus and a weak electromagnetic wave," Phys. Rev. A 100, 52502 (2019).

[55] S. P. Roshchupkin, N. R. Larin, and V. V. Dubov, "Resonant photoproduction of ultrarelativistic electron-positron pairs on a nucleus in moderate and strong monochromatic light fields," Physical Review D 104, 116011 (2021).

[56] S. P. Roshchupkin, A. V. Dubov, V. V. Dubov, and S. S. Starodub, "Fundamental physical features of resonant spontaneous bremsstrahlung radiation of ultrarelativistic electrons on nuclei in strong laser fields," New Journal of Physics 24, 13020 (2022).

[57] A. Hartin, "Second order QED processes in an intense electromagnetic field," arXiv:1701.02906 (2017).

[58] S.P. Roshchupkin, S.S. Starodub, "The effect of generation of narrow ultrarelativistic beams of positrons (electrons) in the process of resonant photoproduction of pairs on nuclei in a strong electromagnetic field", Laser Phys. Lett. 19, 115301 (2022).

[59] S.P. Roshchupkin, A.V. Dubov, and S.S. Starodub, "The possibility of creating narrow beams of high-energy gamma quanta in the process of resonant spontaneous bremsstrahlung radiation of ultrarelativistic electrons on nuclei in strong electromagnetic fields", Phys. Scr. 97, 105302 (2022).

[60] S.P. Roshchupkin, V.D. Serov, V.V. Dubov, "Generation of Narrow Beams of Ultrarelativistic Positrons (Electrons) in the Breit–Wheeler Resonant Process Modified by the Field of a Strong Electromagnetic Wave", Photonics, 10, 949 (2023).

[61] A. Gonoskov, T. Blackburn, M. Marklund, and S. Bulanov, "Charged particle motion and radiation in strong electromagnetic fields", Rev. Mod. Phys., 94, 045001 (2022).

[62] T. G. Blackburn, "Radiation reaction in electron–beam interactions with high-intensity lasers", Rev. Mod. Plasma Phys. 4, 5 (2020).

[63] Y. I. Salamin, S. Hu, K. Z. Hatsagortsyan, and C. H. Keitel, "Relativistic high-power laser–matter interactions", Phys. Rep. 427, 41 (2006).

[64] F. Cajiao V´elez, J. Kamiński, and K. Krajewska, "Electron Scattering Processes in Non-Monochromatic and Relativistically Intense Laser Fields", Atoms 7, 34 (2019).

[65] V. I. Ritus, "Quantum effects of the interaction of elementary particles with an intense electromagnetic field", J. Sov. Laser Res. 6, 497 (1985).

[66] P. Zhang, S. S. Bulanov, D. Seipt, A. V. Arefiev, and A. G. R. Thomas, "Relativistic plasma physics in supercritical fields", Phys. Plasmas 27, 050601 (2020).

[67] R. Ruffini, G. Vereshchagin, and S.-S. Xue, "Electron–positron pairs in physics and astrophysics: From heavy nuclei to black holes", Phys. Rep. 487, 1 (2010).

[68] E. Lötstedt, U. D. Jentschura, and C. H. Keitel, "Coulomb-field-induced conversion of a high-energy photon into a pair assisted by a counterpropagating laser beam", New J. Phys. 11, 013054 (2009).

[69] A. Di Piazza, E. Lötstedt, A. I. Milstein, and C. H. Keitel, "Effect of a strong laser field on electron-positron photoproduction by relativistic nuclei", Phys. Rev. A 81, 062122 (2010).

[70] S. Augustin and C. Müller, "Nonperturbative Bethe–Heitler pair creation in combined high- and low-frequency laser fields", Phys. Lett. B 737, 114 (2014).

[71] C. Müller, A. B. Voitkiv, and N. Grün, "Differential rates for multiphoton pair production by an ultrarelativistic nucleus colliding with an intense laser beam", Phys. Rev. A 67, 063407 (2003).

[72] K. Krajewska and J. Z. Kamiński, "Recoil effects in multiphoton electron-positron pair creation", Phys. Rev. A 82, 013420 (2010).

[73] B. Hafizi, D. F. Gordon, and D. Kaganovich, "Pair Creation with Strong Laser Fields, Compton Scale X Rays, and Heavy Nuclei", Phys. Rev. Lett. 122, 233201 (2019).

[74] A. A. Lebed', "Electron-nucleus scattering at small angles in the field of a pulsed laser wave", Laser Phys. Lett. 13, 045401 (2016).

[75] K. Krajewska and J. Z. Kami´nski, "Breit-Wheeler process in intense short laser pulses", Phys. Rev. A 86, 052104 (2012).

[76] A. Di Piazza and T. Pătuleanu, "Electron mass shift in an intense plane wave", Phys. Rev. D 104, 076003 (2021).

[77] A. Ilderton, Phys. "Trident Pair Production in Strong Laser Pulses", Rev. Lett. 106, 020404 (2011).

[78] A. Hartin, "Strong field QED in lepton colliders and electron/laser interactions", International Journal of Modern Physics A33, 1830011 (2018).

[79] A. A. Mironov, S. Meuren, and A. M. Fedotov, "Resummation of QED radiative corrections in a strong constant crossed field", Phys. Rev. D 102, 053005 (2020).

[80] D. M. Volkov, "On a class of solutions of the Dirac equation", Z. Phys. 94, 250 (1935).

[81] H. Wang, M. Zhong, L.F. Gan, "Orthonormality of Volkov Solutions and the Sufficient Condition", Commun. Theor. Phys. 71, 1179 (2019).

[82] J. Schwinger, "On Gauge Invariance and Vacuum Polarization", Phys. Rev. 82, 664 (1951).

[83] L.S. Brown, T.W.B. Kibble, "Interaction of Intense Laser Beams with Electrons", Phys. Rev. 133, A705 (1964).

[84] G. Breit, E. Wigner, "Capture of Slow Neutrons", Phys. Rev. 49, 519 (1936).

[85] V. B. Berestetskii, E. M. Lifshitz, and L. P. Pitaevskii, Quantum Electrodynamics: Volume 4 (Butterworth-Heinemann, 1982) google-Books-ID: YlwKR5JNWDgC.

[86] G. Breit and J. A. Wheeler, "Collision of two light quanta," Physical Review 46, 1087–1091 (1934).

[87] E. Gerstner, "Lasing at the limit," Nature Physics 6, 638–638 (2010).

[88] Z.-L. Deng, Z.-F. Gao, X.-D. Li, Y. Shao, "On the Formation of PSR J1640+2224: A Neutron Star Born Massive?", Astrophysical Journal 892, 4 (2020).

[89] Z.-L. Deng, Z.-F. Gao, X.-D. Li, Y. Shao, "Evolution of LMXBs under Different Magnetic Braking Prescriptions", Astrophysical Journal 909, 174 (2021).

[90] Z.-F. Gao, N. Wang, H. Shan, X.-D. Li, W. Wang, "The Dipole Magnetic Field and Spin-down Evolutions of the High Braking Index Pulsar PSR J1640–4631", Astrophysical Journal 849, 19 (2017).

[91] H. Wang, Z.-F. Gao, H.-Y. Jia, N. Wang, X.-D. Li, "Estimation of Electrical Conductivity and Magnetization Parameter of Neutron Star Crusts and Applied to the High-Braking-Index Pulsar PSR J1640-4631", Universe. 6, 63 (2020).

[92] Z.-F. Gao, X.-D. Li, N. Wang, J.P. Yuan, P. Wang, Q.H. Peng, Y.J. Du, "Constraining the braking indices of magnetars", Mon. Notices Royal Astron. Soc. 456, 55 (2016).

[93] F.-Z. Yan, Z.-F. Gao, W.-S. Yang, A.-J. Dong, "Explaining high braking indices of magnetars SGR 0501+4516 and 1E 2259+586 using the double magnetic-dipole model", Astron. Nachrichten. 342, 249 (2021).